\documentclass{article}
\usepackage{graphicx,graphics}
\usepackage{amssymb,amsmath,cite}
\newcommand{\degr}{\mbox{$^{\circ}\,$}}

\newcommand{\grad}{$^o$}
\newcommand{\fsi}{\emph{FSI }}

\begin{document}

\title{Study of the $\Lambda p$ Interaction Close to the $\Sigma^+n$\\
and $\Sigma^0p$ Thresholds}

\author{H.~Machner,$^1$ J.~Haidenbauer,$^{2,3}$ F.~Hinterberger,$^4$\\ A.~Magiera,$^5$, J.A.~Niskanen,$^6$ J.~Ritman,$^2$ R.~Siudak,$^7$
\\ \\
$^1$Fachbereich Physik, Universit\"{a}t Duisburg-Essen,\\ Duisburg, Germany\\
$^2$Institut f\"{u}r Kernphysik and J\"{u}lich Centre for Hadron Physics, \\ Forschungszentrum J\"{u}lich, J\"{u}lich, Germany\\
$^3$Institute for Advanced Simulation, \\Forschungszentrum J\"{u}lich, J\"{u}lich, Germany\\
$^4$Helmholtz-Institut f\"{u}r Strahlen- und Kernphysik\\ der Universit\"{a}t Bonn, Bonn, Germany\\
$^5$Institute of Physics, Jagellonian University,\\ Krak\'{o}w, Poland\\
$^6$Department of Physical Sciences, University of Helsinki,\\ Helsinki, Finland\\
$^7$Institute of Nuclear Physics, Polish Academy of Sciences,\\ Krak\'{o}w, Poland}
\maketitle
%
%%%%%%%%%%%%%%%%%%%%%%%%%%%%%%%%%%%%%%%%%%%%%%%%%%%%%%%%%%%%%%%%%%%%%%%%
%
\begin{abstract}
The $\Lambda p$ interaction close to the $\Sigma N$ threshold is considered.
Specifically, the pronounced structure seen in production reactions like
$K^-d \to \pi^- \Lambda p$ and $pp\to K^+ \Lambda p$ around the $\Sigma N$
threshold is analyzed. Modern interaction models of the coupled
$\Lambda N - \Sigma N$ systems generate such a structure either due to
the presence of a (deuteron-like) unstable bound state or of an inelastic
virtual state.
A determination of the position of the prominent peak as observed in
various experiments for the two aforementioned reactions
leads to values that agree quite well with each other.
Furthermore, the deduced mean value of $2128.7\pm 0.3$ MeV for the
peak position coincides practically with the threshold energy of the
$\Sigma^+ n$ channel. This supports the interpretation of the structure
as a genuine cusp, signaling an inelastic virtual state in the
$^3S_1-^3D_1$ partial wave of the $\Sigma N$ isospin 1/2 channel.
There is also evidence for a second peak (or shoulder)
in the data sets considered which appears at roughly 10-15 MeV above
the $\Sigma N$ threshold. However, its concrete position varies significantly
from data set to data set and, thus, a theoretical interpretation is difficult.
\end{abstract}
%
%%%%%%%%%%%%%%%%%%%%%%%%%%%%%%%%%%%%%%%%%%%%%%%%%%%%%%%%%%%%%%%%%%%%%%%%
%
\section{Introduction}\label{sec:Intro}
\setcounter{equation}{0}
Since the discovery of strangeness, the hyperon--nucleon ($YN$)
interaction has been of fundamental interest both theoretically \cite{Dalitz58}
as well as experimentally \cite{Alexander61}. First, its knowledge is important for the
general understanding of the structure of hadrons and their constituents. Further, it is
also needed to explain the spectra of hypernuclei, which, in return,  also convey
information on these interactions especially at small relative energies. Unfortunately,
since hyperons are short lived, this energy region is practically inaccessible by hyperon beams.
Thus, possible bound states are indispensable as the source of information. In this report we
will concentrate on the
$\Lambda p$ interaction at energies close to the threshold of the $\Sigma N$ channels.
These are at 2128.94 MeV (for $\Sigma^+n$) and at 2130.9 MeV (for $\Sigma^0 p$).
In this region experimental data for elastic scattering,
\begin{equation}
\Lambda p\to \Lambda p \ ,
\end{equation} indicate an enhancement in the $\Lambda p$ cross section
as we will discuss in the next sections.
However, the dynamical origin of this enhancement remains unclear so far.
It could be a cusp structure due to (and at) the opening of the $\Sigma N$ threshold
and then would be a signal for an inelastic virtual state
(we follow here the nomenclature used and explained in Ref.~\cite{Badalyan82})
or due to a bound $\Sigma^0p$ or $\Sigma^+n$ state, i. e. a deuteron-like but unstable bound state.
In the latter case the peak of the cross section has to be below the $\Sigma N$ threshold.
In principle, it could also be a $\Lambda p$ resonance above the $\Sigma N$ threshold.

Data from elastic scattering experiments that cover this range exist in the literature,
but so far the momentum resolution of the $\Lambda$ beams has been insufficient to draw
firm conclusions.
A much more promising avenue is offered by the study of final state interactions (\emph{FSI}).
Assuming relative weakness of the pion interaction one possibility is the strangeness exchange
reaction
\begin{equation}\label{equ:K_induced} K^-d\to\pi^-\Lambda p.
\end{equation}
Also strangeness production processes like
\begin{gather}
\pi^+d\to K^+\Lambda p\label{gat:pi-induced}\\ pp\to K^+\Lambda p\label{gat:reac3}
\end{gather}
should contain basically the same information.

In this work we will concentrate on the two
reactions (\ref{equ:K_induced}) and (\ref{gat:reac3}).
With regard to the former reaction, evidence for an enhancement in the
$\Lambda p$ cross section near the $\Sigma N$ threshold has been already found
in the late 1960s and confirmed in later experiments
\cite{Tan69, Cline68, Braun77, Sims71, Alexander68, Eastwood71, Pigot85}.
We review those data and we also re-analyze them with the aim to determine
accurately the position of this enhancement.
Experimental information about the reaction \ref{gat:reac3}, in the region of the $\Sigma N$
threshold, has become available much more recently and clear evidence for
the presence of an enhancement at that threshold is only emerging right now.
Here we perform an analysis of data from the inclusive measurements of the reaction
$pp\to K^+ X$ performed at Saclay~\cite{Siebert94} and in J\"{u}lich \cite{Budzanowski10a},
respectively, and attempt to
extract the enhancement and its position from those experiments too.
This sort of analysis suffers from lacking precise $pp$ {\it exclusive} data
to rely on. However, as we will see, the positions determined from the two reactions (\ref{equ:K_induced})
and (\ref{gat:reac3}) agree with each other, and they also coincide with the
opening of the $\Sigma^+ n$ channel within the error bars of our analysis.

We also study an additional peak (or shoulder) that is present in basically all the
aforementioned measurements and located a few MeV above the $\Sigma N$ threshold.
However, in this case it turns out that there are sizable variations of its position
between reaction (\ref{equ:K_induced}) and (\ref{gat:reac3}), but even between
measurements of one and the same reaction. Thus, the physical significance of that
peak remains unclear to some extent.

The paper is organized as follows. In the next section we take a look at
the status of the results from elastic $\Lambda p$ scattering. We discuss also
the behavior of the $\Lambda p$ cross section around the $\Sigma N$ threshold
as predicted by various $YN$ interaction models from the literature.
In the two subsequent sections the data for the two reactions (\ref{equ:K_induced})
and (\ref{gat:reac3}) are discussed.
In section 5 the peak structures found are analyzed. In particular, we determine
their position as seen in the various measurements for the two reactions in
question. The paper ends with a summary.

\section{Elastic ${\Lambda p}$ Scattering}
\label{sec:Elastic-Lambda-p-Scattering}

Before discussing the situation for $\Lambda N$ scattering let us briefly recall
some well-known features of coupled-channels dynamics \cite{Newton66,Badalyan82}.
Conservation of flux and the associated unitarity of the $S$-matrix necessarily
imply anomalies at the opening of new thresholds \cite{Newton66}.
Specifically, at an $S$-wave threshold the cross section of the ``old'' channel as a
function of the energy will, in general,
have infinite slopes at such a threshold. The resulting structures are usually
called cusps or rounded steps, depending on their specific shape
\cite{Newton66,Badalyan82}. Whether these structures remain primarily of academic
interest or manifest themselves via large, experimentally observable effects depends
strongly on the strengths of the interactions in the coupled channels. In particular,
pronounced threshold phenomena always go along with near-by poles in the scattering
amplitudes of the involved channels that are associated with (inelastic) virtual
states or (unstable) bound states \cite{Badalyan82,Miyagawa99}.

Modern meson-exchange models of the $YN$ interaction such as the J\"{u}lich
\cite{Haidenbauer05, Holzenkamp89} or Nijmegen potentials \cite{Rijken99, Maessen89}
are derived under the assumption of (broken) SU(3) symmetry.
This symmetry implies, that the strongly attractive forces that yield the deuteron
bound state (in the $^3S_1-{^3D_1}$ partial wave) and a virtual
state in the $^1S_0$ partial wave in case of the $NN$ system will likewise act
in the strangeness $S=-1$ sector
(see, e.g., Refs.~\cite{Polinder06,Polinder07} for details on the SU(3) relations).
Specifically, there is a strong coupling between the $\Lambda N$ and $\Sigma N$ systems.
It is caused by the long-ranged tensor force provided by pion exchange and
boosted by the fact that the thresholds of the two channels are only
separated by 77 MeV.
Therefore, it is not surprising that practically all $YN$ interactions
that fit the data and include explicitly the coupling between the
$\Lambda N$ and $\Sigma N$ channels predict also sizeable threshold effects.

\begin{figure}[t]
\begin{center}
\includegraphics[width=0.46\textwidth]{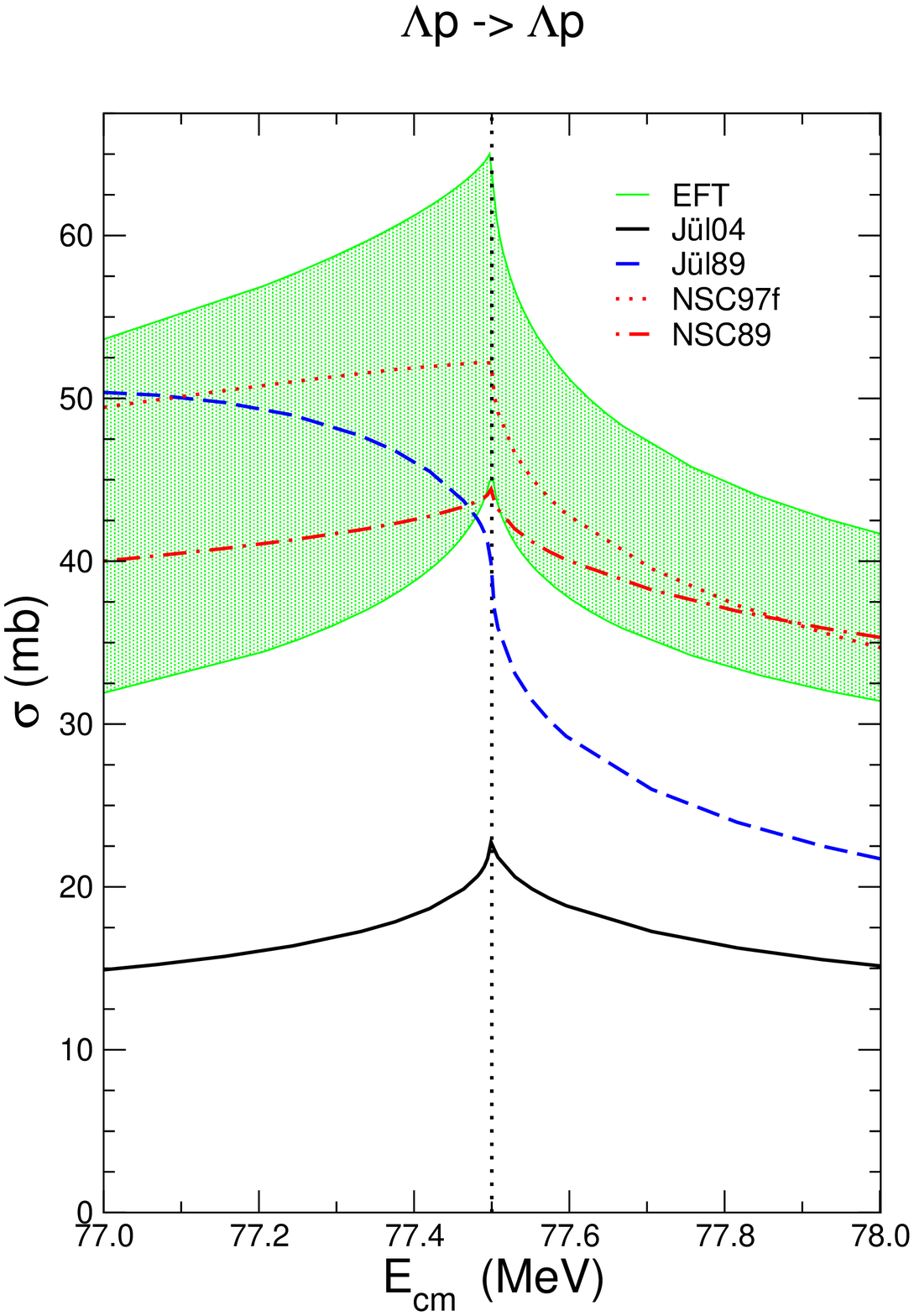}
\includegraphics[width=0.47\textwidth]{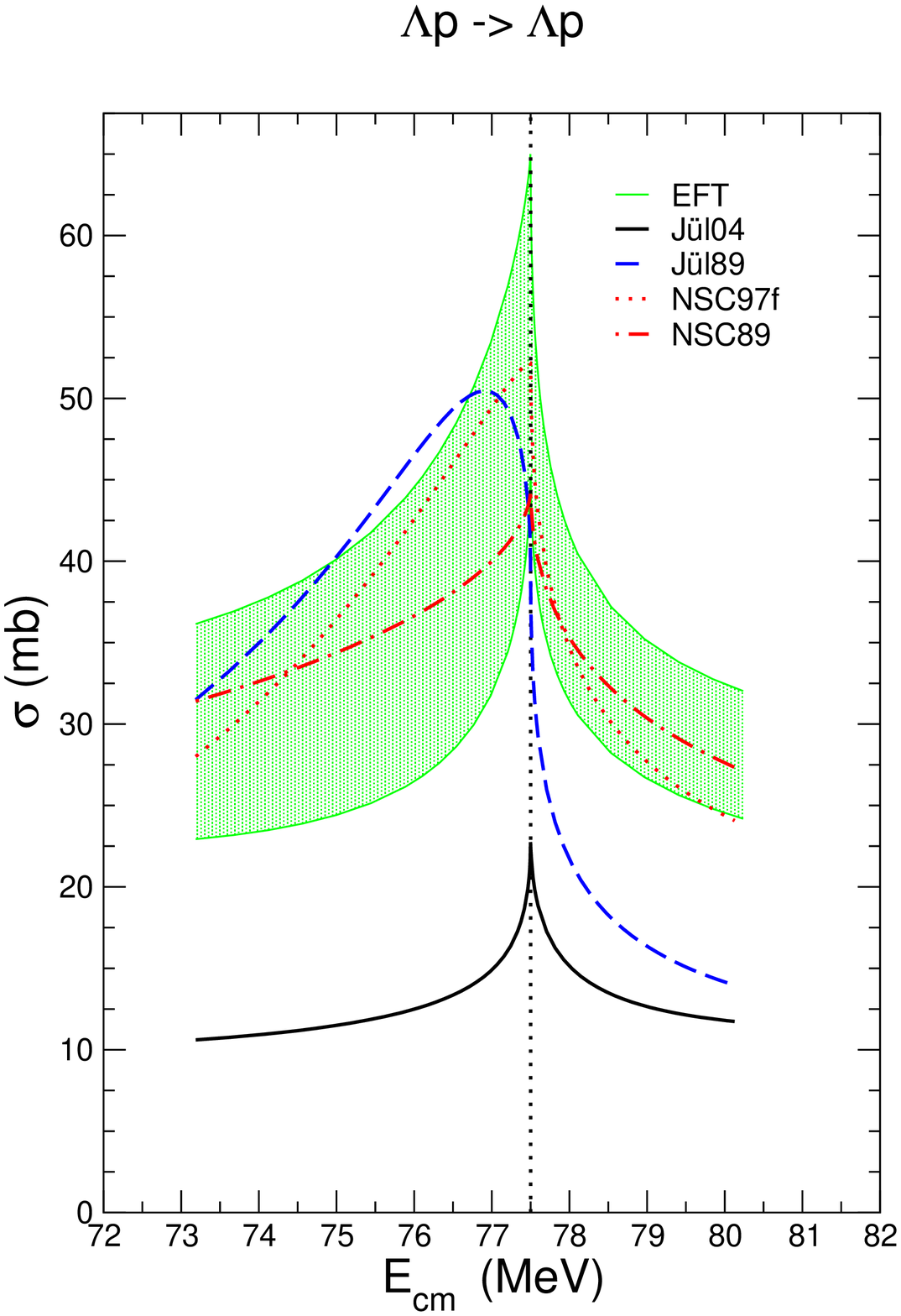}
\caption{Elastic $\Lambda N$ cross section as function of the center-of-mass energy.
J\"{u}l04 and J\"{u}l89 are results for the J\"{u}lich potentials
published in Refs. \cite{Haidenbauer05} and \cite{Holzenkamp89}, respectively.
NSC97f and NSC89 refer to results of the corresponding Nijmegen soft-core
potentials \cite{Rijken99} and \cite{Maessen89}. Results obtained at leading-order
chiral EFT \cite{Polinder06,Polinder07} are indicated by the grey band.
The dashed line is the threshold for the $\Lambda p\to \Sigma N$ transition.
The left part has an expanded energy scale.
}
\label{Fig:Phases}
\end{center}
\end{figure}
A closer inspection of the results for published interaction models reveals
that, in general, they can be grouped into two categories. In one case the
predicted $\Lambda N$ cross
section around the $\Sigma N$ threshold is in the order of 40 to 50 mb and, more
characteristic, roughly a factor four larger than a few MeV away from the threshold.
The cross section is so large because one of the eigenphases of the tensor-coupled
$^3S_1-^3D_1$ partial wave (in most cases the $^3D_1$) passes through 90\degr right
below the $\Sigma N$ threshold. Due to the latter aspect, the peak of the
cross section is actually not at the $\Sigma N$ threshold but slightly below.
Thus, in this case the $\Lambda N$ cross section exhibits a typical
resonance-like behavior. Moreover, no cusp appears at the actual
$\Sigma N$ threshold, only a rounded step \cite{Newton66,Badalyan82}.
However, since the peak of the cross section occurs so close to the
$\Sigma N$ threshold -- often the separation is less than an MeV --
it is usually impossible to recognize the above features in the published
results due to the scale used for the figures!
Pole searches performed for the amplitudes produced by the potential models in
question found that these are located in the second quadrant of the complex plane
of the relative momentum in the $\Sigma N$ channel \cite{Miyagawa99}.
Thus, these $YN$ interactions are characterized by the presence of an unstable
bound state, i.e. a deuteron-like $\Sigma N$ state \cite{Badalyan82}.
The Nijmegen potentials NSC97f \cite{Rijken99} and NF \cite{Nagels79}  but also
the original J\"{u}lich $YN$ interaction \cite{Holzenkamp89} belong to this category.
In the case of an unstable bound state it is also possible that the $^3S_1$ eigenphase
passes through 90\degr (instead of the $^3D_1$). Interestingly, such a scenario is
seldom realized. In fact, we are aware of only one meson-exchange $YN$ potential
where this happens, namely the Nijmegen ESC04 interaction \cite{Rijken06}.
In case of the interactions considered in \cite{Toker81} it is also the $^3S_1$
that passes through 90\degr. But since these potentials were intended for application
in Faddeev-type calculations for simplicity reasons only $S$-waves were taken
into account. One of the interactions considered in \cite{Toker81} has the
rather unique feature that the predicted $^3S_1$ phase passes through 90\degr
slightly above the $\Sigma N$ threshold. As far as we can see, this does not
happen for any of the meson-exchange $YN$ potentials whose phase shifts are
documented in the literature.

The second category of $YN$ potentials produces a peak in the $\Lambda N$
cross section precisely at the $\Sigma N$ threshold.
Thus, now we do observe a genuine threshold cusp.
Here none of the relevant eigenphases passes through 90\degr.
In general the $\Lambda N$ cross section at the $\Sigma N$ threshold is roughly a factor two larger
than at a few MeV away from the threshold. Usually, the peak values are around 20 mb, but can
still reach up to 40 mb.
The poles for this kind of potentials are located in the third quadrant of the
complex plane of the relative momentum in the $\Sigma N$ channel \cite{Miyagawa99}.
They are an indication for the presence of inelastic virtual states \cite{Badalyan82},
i.e. the analog of the virtual state in the $NN$ $^1S_0$ partial wave.
The Nijmegen potentials ND \cite{Nagels77}, NSC89 \cite{Maessen89} and ESC08 \cite{Rijken2010},
and also the recent J\"{u}lich $YN$ interaction \cite{Haidenbauer05} belong to this
category.
Also a $YN$ interaction derived at leading-order in chiral effective field theory (EFT)
\cite{Polinder06,Polinder07} predicts such a behavior.

In order to illustrate the statements made above, $\Lambda p$ cross sections
for the Nijmegen soft-core potentials NSC97f \cite{Rijken99} and
NSC89 \cite{Maessen89}, and for the J\"{u}lich potentials
from 2005 \cite{Haidenbauer05} and 1989 \cite{Holzenkamp89} are shown
in Fig.~\ref{Fig:Phases} for energies around the $\Sigma N$ threshold for two
different energy scales. Results obtained at leading-order
chiral EFT \cite{Polinder06,Polinder07} are indicated by the grey band.
One can see that the cross sections predicted by those $YN$ potentials
indeed exhibit different features at the $\Sigma N$ threshold.
As said above the Nijmegen NSC97f and the J\"{u}lich 1989 one-boson
exchange models produce a deuteron-like unstable bound state in the $\Sigma N$ channel.
In both cases the $^3D_1$ $\Lambda N$ phase shift shows a resonance-like
behavior and crosses 90\degr slightly below the $\Sigma N$ threshold
\cite{Haidenbauer05,Polinder06}.
The rounded step in the cross section of the J\"{u}l89 potential
is clearly visible in the left figure (with magnified scale). The structure
produced by the NSC97f potential is similar. However, since in this
case (in our calculation) the $^3D_1$ phase shift crosses 90\degr at a
mere 20 keV below the nominal $\Sigma N$ threshold a further increase in
the scale would be required to see that.
The J\"{u}l04 and the NSC89 models and the EFT interaction support an
inelastic virtual state rather than a bound state and, consequently, a
genuine cusp structure appears in the cross section.

\begin{figure}[h]
\begin{center}
\includegraphics[width=0.8\textwidth]{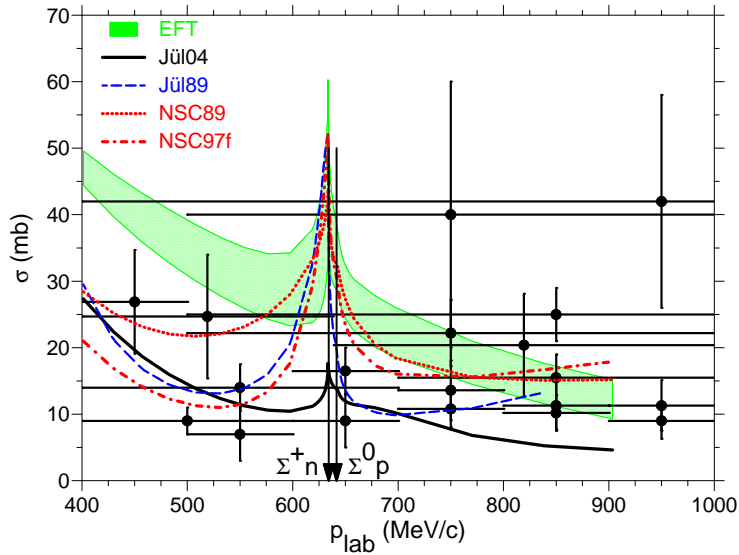}
\caption{Cross sections for elastic $\Lambda p$ scattering.
Data are from \cite{Hauptman77, Kadyk71, Alexander61, Cline67, Crawford59, Charlton70}.
The curves are results from $YN$ models,
namely from the Nijmegen $YN$ soft-core potentials NSC97f \cite{Rijken99} (dashed curve)
and NSC89 \cite{Maessen89} (dash-dotted curve),
and from the J\"{u}lich one-boson-exchange models \cite{Haidenbauer05} (solid curve).
and \cite{Holzenkamp89} (dashed curve). Results obtained at leading-order
chiral EFT \cite{Polinder06,Polinder07} are indicated by the grey band.
The thresholds for the reactions $\Lambda p\to \Sigma N$ are indicated by arrows.}
\label{Fig:elastic}
\end{center}
\end{figure}

Note that that the results in Fig.~\ref{Fig:Phases} are
calculated in isospin basis using isospin-averaged $\Sigma$ and nucleon
masses. In particular, the $\Lambda N$ momentum and the $\Sigma N$
threshold are evaluated for $m_N=[m_p+m_n]/2$ and
$m_\Sigma=[m_{\Sigma^+}+m_{\Sigma^0}+m_{\Sigma^-}]/3$, respectively.

Experimentally, in principle, the $\Lambda p$ interaction could be studied
in the range of interest, i. e. in the vicinity of the $\Sigma N$
thresholds by elastic scattering. In Fig. \ref{Fig:elastic} all
cross sections \cite{Hauptman77, Kadyk71, Alexander61, Cline67,
Crawford59, Charlton70} -- to the best of our knowledge -- for elastic
scattering in a momentum range from 400 MeV/c to 1 GeV/c are collected.
In addition to data we include the model calculations for the Nijmegen soft-core
potentials NSC97f \cite{Rijken99} and NSC89 \cite{Maessen89}, the
J\"{u}lich meson-exchange models J\"{u}l05\cite{Haidenbauer05} and J\"{u}l89\cite{Holzenkamp89},
and the leading-order chiral EFT interaction \cite{Polinder06,Polinder07}.

Here the computation of the cross section was done in
particle basis so that the $\Sigma^+ n$ and $\Sigma^0 p$ thresholds could be
correctly implemented.
Partial waves up to $L\leq 2$ have been taken into account.
Note that the agreement between data and calculations at low energies
(not shown here) is of similar quality for all models.

Obviously, the data on elastic scattering
are insufficient in quality to allow to discriminate between the
different scenarios. It is therefore highly desirable to have
additional data of high quality.

\section{The reaction $ {K^-d\to \pi^-\Lambda p}$.}
\label{sec:reactio-K-}

One possibility to study the $\Lambda p$ interaction is via the \fsi in
reactions like $K^-d\to \pi^-\Lambda p$.
In elastic scattering the contribution from the spin-triplet
waves to the cross section can be expected to be the larger one due
to its statistical weight. For the above reaction we expect likewise
the main contribution to come from
spin-triplet states and, specifically, from the $^3S_1$-wave,
but due to different reason: the deuteron is already
present in the initial state. For kaon absorption at rest the final
baryonic state has to have predominantly the quantum numbers of the deuteron.
At lowest order the reaction should be dominated by $K^-n\to\pi^-\Lambda$
(quasi-elastic) scattering leaving the spin-space part of the baryon state unchanged.
\begin{figure}
\begin{center}
\includegraphics[width=0.8\textwidth]{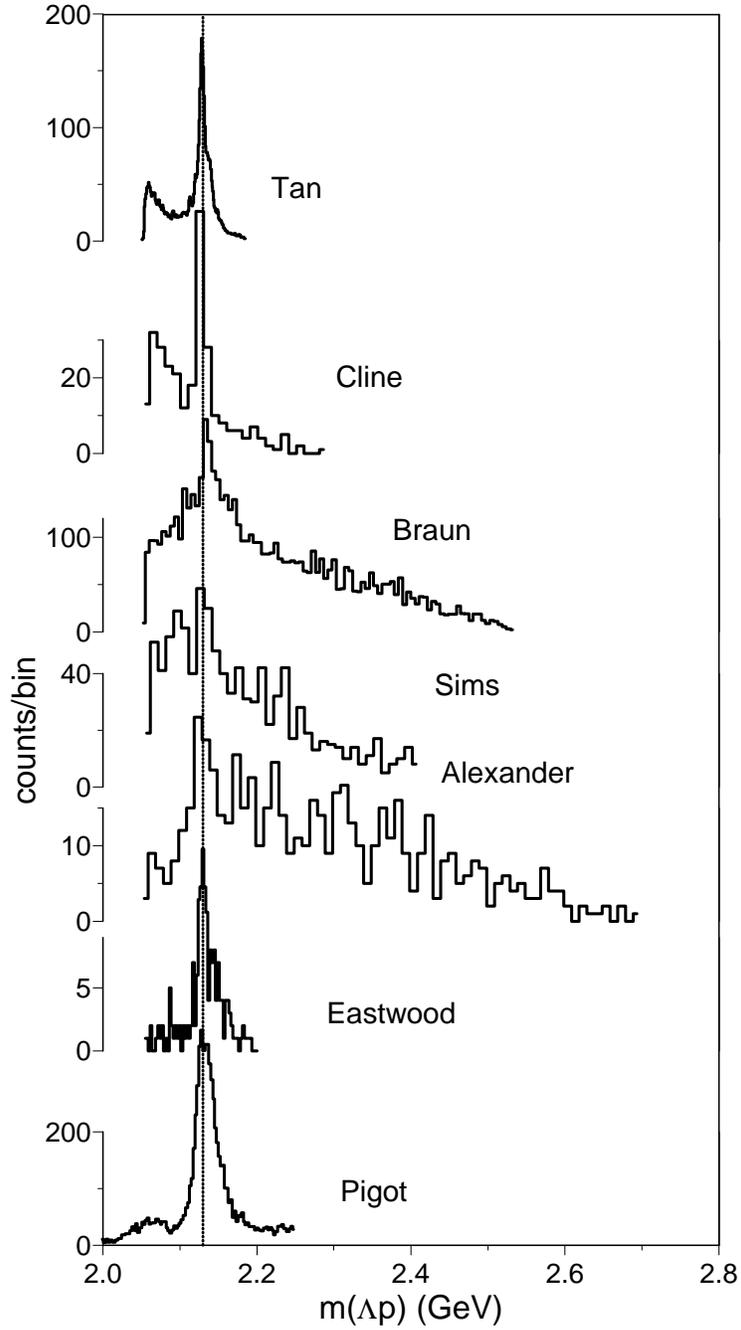}
\caption{Invariant mass spectra for the reaction $K^-d\to \pi^-\Lambda p$.
The data are from Tan \cite{Tan69}, Cline \textit{et al.} \cite{Cline68},
Braun \textit{et al.} \cite{Braun77}, Sims  \textit{et al.}  \cite{Sims71},
Alexander \textit{et al.} \cite{Alexander68}, Eastwood \textit{et al.} \cite{Eastwood71},
and  Pigot \textit{et al.} \cite{Pigot85}. The dotted line indicates the
averaged $\Sigma N$ mass 2.13 GeV.}
\label{Fig:Pi-Lp}
\end{center}
\end{figure}

\begin{figure}
\begin{center}
\includegraphics[width=0.8\textwidth]{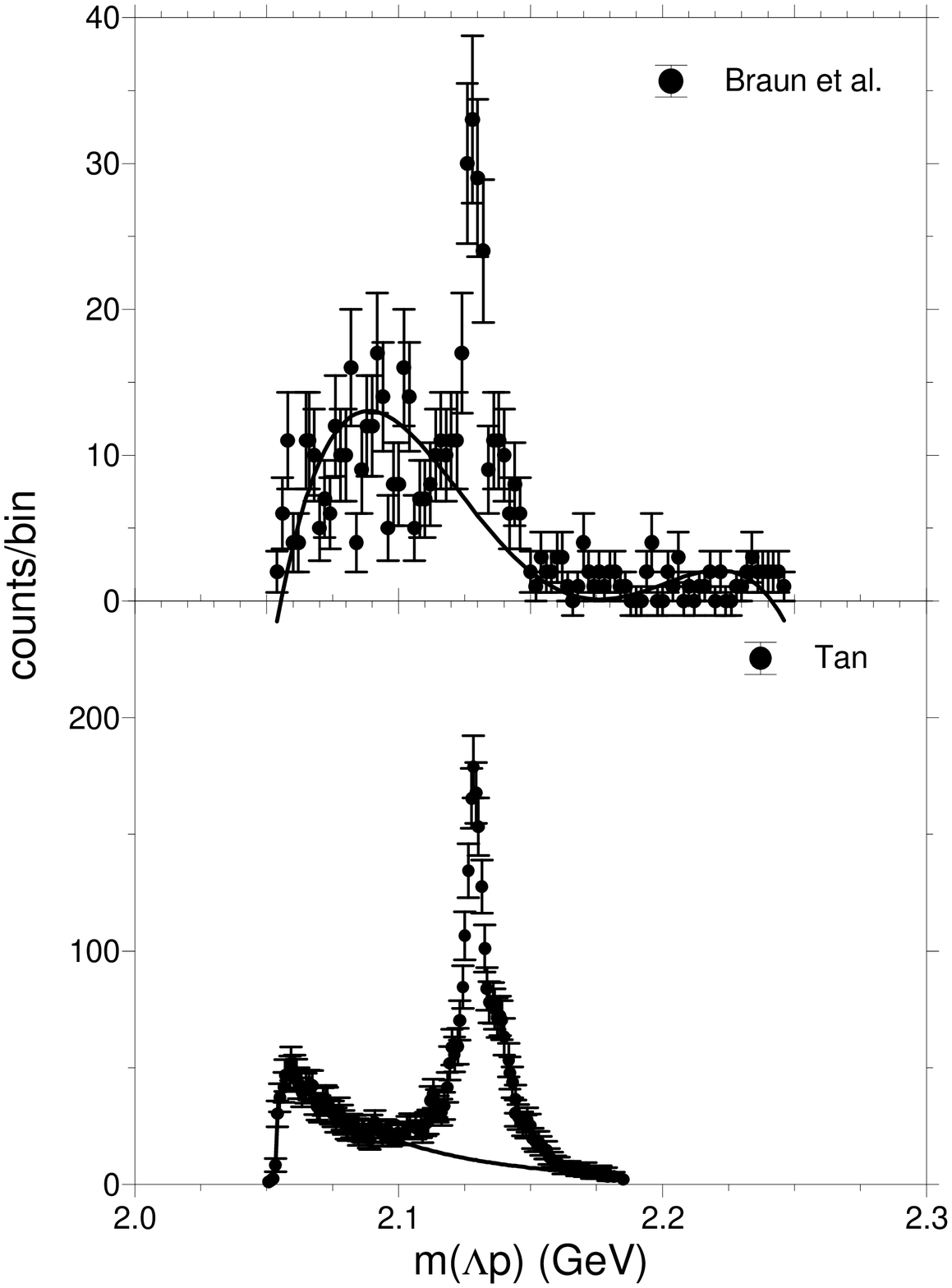}
\caption{The missing mass spectra for the constrained data (see text) from
\cite{Braun77} (upper panel) and \cite{Tan69} (lower panel). The solid curves
show fits to the data where the peak region has been excluded.}
\label{Fig:Braun_Tan}
\end{center}
\end{figure}

Fig.~\ref{Fig:Pi-Lp} shows the $\Lambda p$ spectra in the center-of-mass (c.m.) system. The data are taken from Refs.~\cite{Tan69, Cline68, Braun77, Sims71, Alexander69, Eastwood71, Pigot85}. There is clearly a peak around the $\Sigma N$ mass, i.e. around 2.13 GeV. However, the shape of the peaks as well as the underlying cross section vary from one experiment to the other. One reason for this are different boundary conditions in the experiments and the analysis. We will return to this point later. In order to study this dependence further we compare in more detail some of the spectra. The data from Sims \textit{et al.} \cite{Sims71} are ignored since they have a poor signal to background ratio, i.e. do not allow to study the structure in detail. Similarly, the data from Alexander \textit{et al.} \cite{Alexander69} as well as those from Cline \textit{et al.} \cite{Cline68} with bin widths of 10 MeV and poor statistics are excluded. The latter data have only two bins with 67 counts together in the peak region thus
a width of the peak cannot be extracted. The data from Eastwood \textit{et al.} \cite{Eastwood71}
on the other hand contain some more counts (130) but in thirteen bins. We therefore kept these data.
First we inspect the spectra from Tan \cite{Tan69} and Braun et al. \cite{Braun77}. These are data sets with
rather large statistics. The smooth yield below the peaks contains, in addition to the direct reaction
(\ref{equ:K_induced}), contributions from the reaction with a heavier intermediate hyperon:
$K^-d\to \Sigma^-(1385)p$ which decays in a second step
as $\Sigma^-(1385)\to \Lambda\pi^-$.
In order to avoid any phenomenological modelling and possible ambiguities associated with that
we simply subtract some yield below the peak by fitting polynomials to these yields.  We will discuss
this further in section \ref{sub:higher-Mass-structu}. The result is depicted in Fig. \ref{Fig:Braun_Tan}.
Then this fitted cross section is subtracted from the experimental cross section so that only the peak structure remains. The peak in Tan's paper \cite{Tan69} shows a shoulder on the heavy mass side. In the case of the unconstrained data from Braun et al. \cite{Braun77} the structure looks quite different. There is much more yield to the high-mass side than in Tan's data. At this point we have to elucidate that this comparison is made between different things. The data from Braun \textit{et al.} have no constraints while in case of Tan, counts with proton momenta below 75 MeV/c have been cut. Thus, reactions with the proton as a mere spectator $K^-d\to\pi^-\Lambda p_s$, with the two baryons being uncorrelated are excluded. In order to have really both hadrons in the entrance channel participating in the reaction Braun \textit{et al.} introduced two cuts: (i) a threshold in the proton momentum of 150 MeV/c and (ii) the requirement of the angle between the incoming kaon and the outgoing pion $0.9<\cos(K,\pi)<1$. If one takes that into account the peak agrees, with respect to its shape and its position,
to a large extent to the one reported by Tan \cite{Tan69} as is discussed
in Sect.~\ref{sec:Peak-structu-reactio-mechani}.
Note that Eastwood \textit{et al.} \cite{Eastwood71} required the proton momenta to be larger than 170 MeV/c.

While almost all groups relied on bubble chambers, Pigot \textit{et al.} \cite{Pigot85}
applied a magnetic spectrometer to detect the emerging pions.
However, most $\pi^-$'s are not from reaction (\ref{equ:K_induced}) but from beam
decays $K^-\to 2\pi^-\pi^+$. In order to reduce the number of such events the target
cylinder was surrounded by twelve scintillation counters. After the decay of the
$\Lambda$ into two charged particles one has then three charged particles in addition
to the forward going pion. Therefore, the authors studied the data with charged particle
multiplicity $m\ge 2$ in these scintillators. For the width of the peak they give only
upper limits. They also studied the line reversed reaction (\ref{gat:pi-induced}). The
cross section for this reaction is smaller than for the strangeness exchange. Spectator
protons are rarely detected in the set-up since they are stopped in the liquid deuterium
target being 4 cm in diameter. Their final result is 2129$\pm$0.2$\pm$0.2 MeV and
16.7$\pm$1.9$\pm$2 MeV for the position and the width of the peak, respectively. Here we make use
of their data set of the reaction (\ref{equ:K_induced}) taken at a beam
momentum of 1.4 GeV/c and multiplicity $m \ge 3$. For this case the spectrum is rather clean and the statistics is still sufficient -- which is not the case for the other data sets. We then proceed as in the other cases. The final results for the structure at the $\Sigma N$ threshold are given for all cases in Figs. \ref{Fig:0ne-BW} and \ref{Fig:all-peaks}.

\section{The reaction ${pp\to K^+\Lambda p}$}\label{sec:reactio-pp}
Most studies of the reaction $pp\to K^+\Lambda p$ report only total cross sections. Hogan \textit{et. al} \cite{Hogan68} and more recently the COSY-TOF collaboration \cite{TOF06,TOF10} published spectra. However, these experiments although having rather large acceptances  suffer
from insufficient resolution to study a peak in the threshold region. These unfavorable boundary conditions
were overcome with sufficient high resolution by Siebert et al.~\cite{Siebert94}
employing the SPES4 spectrometer at Saclay and, more recently, even
more by the HIRES experiment \cite{Budzanowski10, Budzanowski10a} making use of
the Big Karl spectrometer \cite{Drochner98} at the COoler SYnchrotron COSY J\"{u}lich.
The disadvantage of the experiments from Refs. \cite{Hogan68, Siebert94, Budzanowski10, Budzanowski10a}
is that they are inclusive.
This means that above the $\Sigma N$ threshold also the $\Sigma^0p$ and $\Sigma^+n$ channels
contribute to the experimental cross section. Thus a peak in the experimental spectrum is
more difficult to see because the signal will be distorted by the rising contributions from the $\Sigma N$ channels. Therefore, the following analysis is incomplete and should be revisited upon
future availability of good near-threshold $\Sigma-N$ production cross sections.

Already at first glance there are differences between
reactions (\ref{equ:K_induced}) and (\ref{gat:reac3}). For example,
available data suggest that the peak to background ratio of the structure
around the $\Sigma N$ threshold is significantly smaller for the latter reaction.
Furthermore, the detection of a kaon in the final state is much more
delicate as compared to the detection of a pion, due to its short life time.
One thing in common, after applying the cuts as discussed in the previous
section, is that both baryons participate in the reactions.

Here we will first concentrate on the data from Refs. \cite{Siebert94} and \cite{Budzanowski10a}.
In these experiments the four momentum of the kaon was measured and the missing mass of the $YN$
system was deduced. The missing mass resolution in the Saclay experiment
\cite{Siebert94} varied between 3 to 5 MeV (FWHM) depending on the
angle while the HIRES experiments \cite{Budzanowski10} and
\cite{Budzanowski10a} reported a resolution of 0.8 MeV.

The bubble chamber measurements and the exclusive COSY-TOF data are projections from Dalitz plots and hence they are in the c.m. system. To make the other data comparable to them we transform them into the c.m. system.

\begin{figure}[h]
\begin{center}
\includegraphics[width=0.8\textwidth]{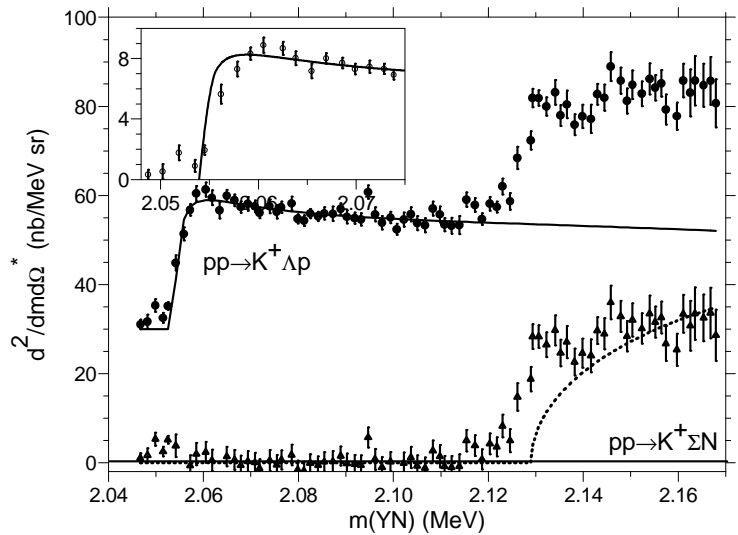}
\caption{Upper curve: The data from SPES 4 (full dots) \cite{Siebert94} at a lab. angle of (10$\pm$2)~\grad. The beam had a momentum of 3.1 GeV/c. The data are transformed into the c.m. system and shifted by 5 nb/MeV sr in height and by 2 MeV down in the mass scale. The solid curve is the fit with Eq. (\ref{equ:PSFSI}) making use of the \fsi parameters from Ref. \cite{Budzanowski10}. The lower curve (full triangles) is the difference between the data in the upper curve and the fit. The short dashed curve is the fit of phase space for $\Sigma^+ n$. The insert shows again the solid curve (without the 5 nb/MeV sr shift and the data without the mass shift of 2 MeV.)  }
\label{Fig:Spes4}
\end{center}
\end{figure}

The authors of Ref. \cite{Siebert94} report about several energy shifts they have applied to their data. We shift the data back by two MeV in order that the steep rise of the cross section agrees with the threshold for the $\Lambda p$ channel. This is demonstrated in Fig. \ref{Fig:Spes4} where the shifted data are compared with calculated cross section
 \begin{equation}\label{equ:PSFSI}
 \frac{d^2\sigma}{dm d\Omega^*}\propto \Phi_3 \cdot F_{FSI}
 \end{equation}
with $\Phi_3$ the three body phase space and $F_{FSI}$ is the enhancement factor for the
final state interaction from Ref. \cite{Budzanowski10}. In order to show the effect of 2 MeV
we have shifted this curve by two MeV up. This curve cannot account for the data points in
the steep rise close to threshold (see insert in Fig. \ref{Fig:Spes4}). The data close to the threshold agree well with FSI modified
phase space. This $\Lambda p$ yield represented by Eq. (\ref{equ:PSFSI}) is then subtracted from
the experimental spectrum.
The remaining yield is then assumed to be due to the contributions from the $pp\to K^+\Sigma^+ n$
and $pp\to K^0\Sigma^+ p$ reactions, and from the structure of interest.
Finally we subtract the cross section of the
two $\Sigma$ channels assuming $K^+\Sigma N$ being phase space distributed.
This is also shown in Fig. \ref{Fig:Spes4}.

The same procedure is applied to the HIRES data.
\begin{figure}[h]
\begin{center}
\includegraphics[width=0.8\textwidth]{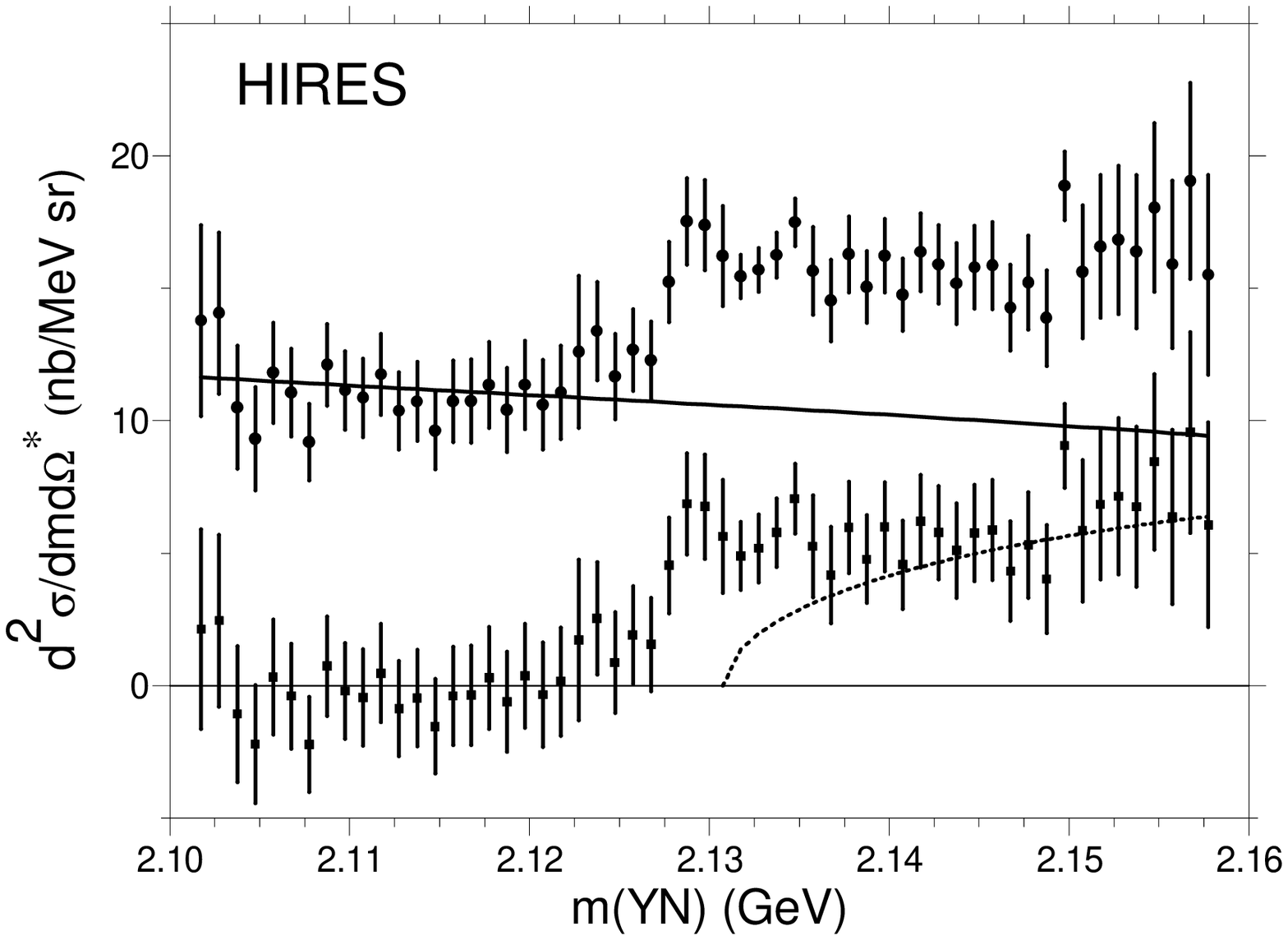}
\caption{Same as Fig. \ref{Fig:Spes4} but for the HIRES data \cite{Budzanowski10a} measured around zero degree and at a beam momentum of 2.87 GeV/c.}
\label{Fig:Hires}
\end{center}
\end{figure}
The different steps are depicted in Fig. \ref{Fig:Hires}. By this method the structure close to the $\Sigma N$ thresholds is extracted.

\begin{figure}[t]
\begin{center}
\includegraphics[width=0.6\textwidth]{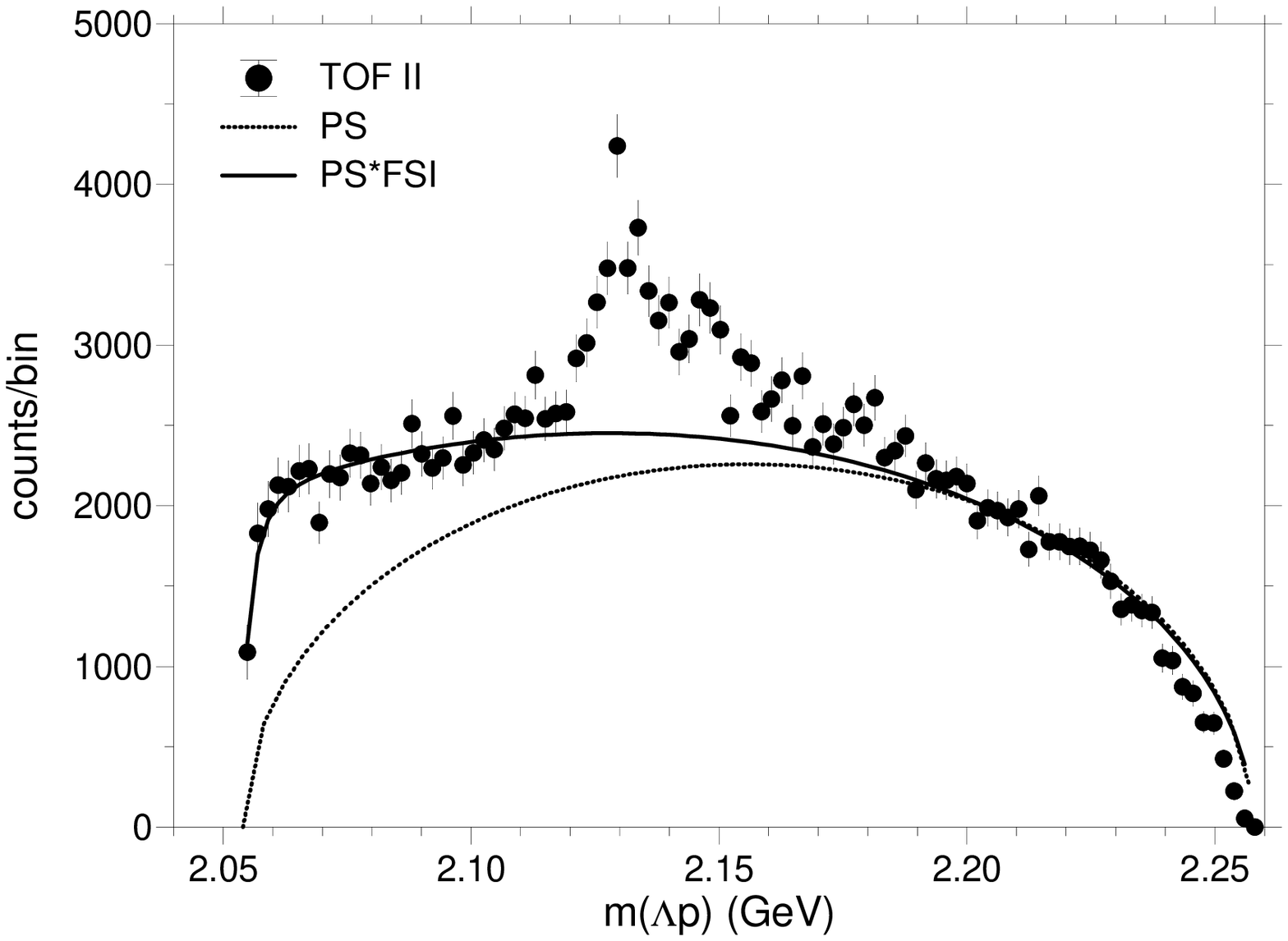}
\caption{The exclusive cross sections for the reaction $pp\to K^+\Lambda p$ at 2.95 GeV/c from Ref. \cite{TOF_12_1, Roeder11}. The curves indicate the fit to the data outside the peak region. The solid curve is the fit with Eq. (\ref{equ:PSFSI}), the dashed curve is the pure phase space part.  }
\label{Fig:Tof}
\end{center}
\end{figure}

The new data from TOF \cite{TOF_12_1} are exclusive. In general such data are superior to inclusive data since subtraction of cross sections due to other reactions is unnecessary. The spectra as function of the $\Lambda p$ mass were obtained by projection of the Dalitz plot onto this axis. The particles detected in the final state were $K^+$, $p$ and the decay $\Lambda\to \pi^-p$.
The first set of data labelled here by TOF~I were taken at a beam momentum of 3.003 GeV/c while the second one was
taken at 2.95 GeV/c.
The resolution of the first set is 6 MeV after a kinematical fit while for the second 2 MeV is reported
due to a new tracking system \cite{Roeder11}.
We proceed as before by fitting Eq. (\ref{equ:PSFSI}) to the data outside the peak region.
This is shown for the TOF~II data set in Fig. \ref{Fig:Tof}.
This procedure accounts very well for the exclusive data. Especially the high mass side is well reproduced by pure phase space.  We can still go a step further and compare the peak regions of the exclusive data with the inclusive data.
\begin{figure}[h]
\begin{center}
\includegraphics[width=0.6\textwidth]{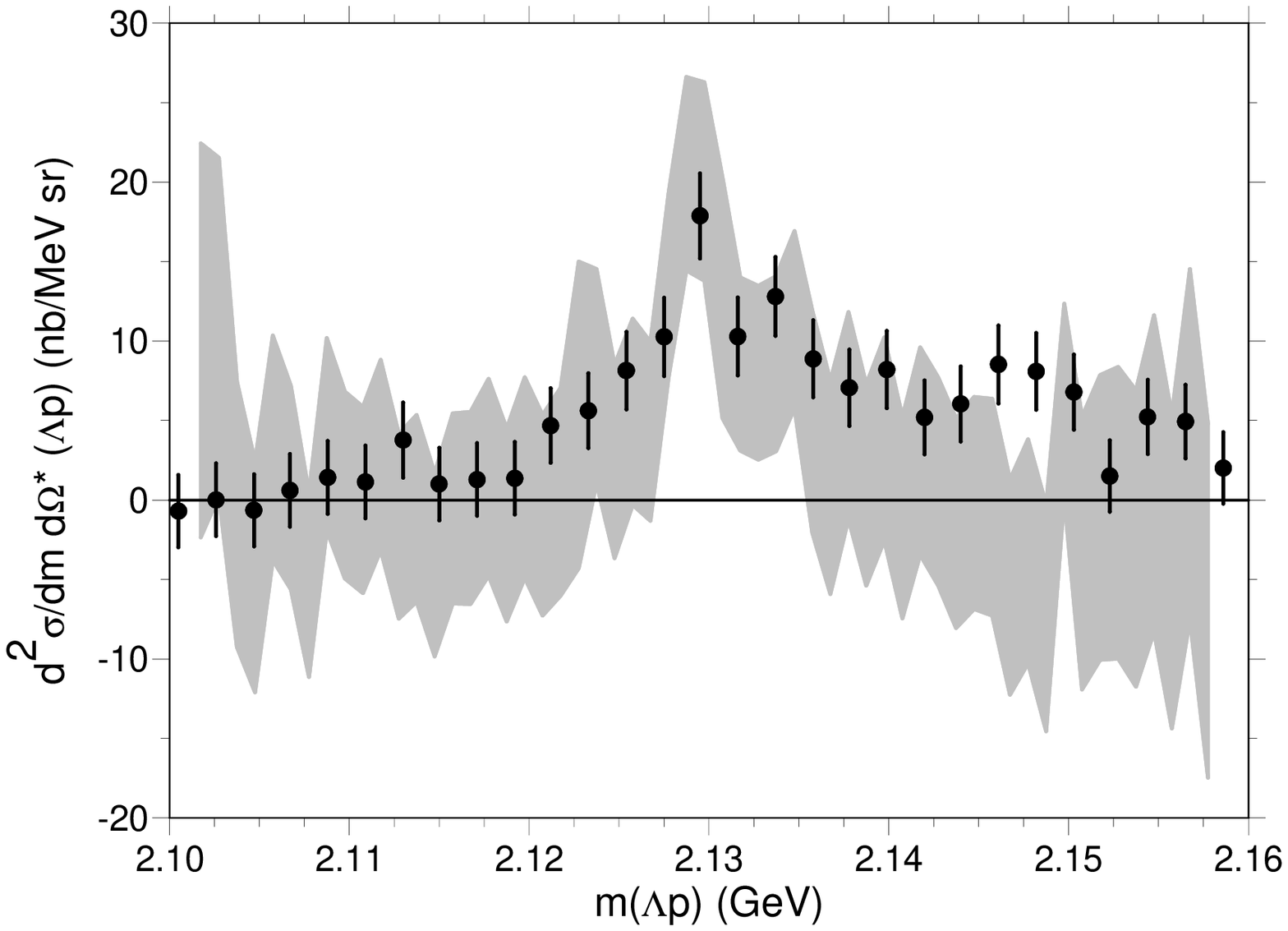}
\caption{Comparison of spectral shapes in the peak region for the inclusive data from HIRES \cite{Budzanowski10a} and TOF \cite{TOF_12_1, Roeder11}. The cross sections $d^2\sigma/dm d\Omega^*$ deduced for the HIRES data are shown as shaded area indicating the error band. The data $d\sigma/dm$ deduced from the TOF~II data set divided by 300 are shown as full circles with error bars. }
\label{Fig:Hires-TOF}
\end{center}
\end{figure}
This comparison is made in Fig. \ref{Fig:Hires-TOF}. There is a remarkable degree of similarity between
the two results although the latter data are taken at a beam momentum 80 MeV/c higher than in the case
of the HIRES data. The similar shape in the right side of the peak - except for $m(\Lambda p)\approx 2.1475$ GeV
- contradicts the statement made in Ref. \cite{Valdau11} that a large fraction of the cross section ascribed
to the $\Sigma N$ production might be $\Lambda p$ production.

At this point it might be appropriate to comment on that work further. First the authors state  that "it is further hypothesised (in Ref. \cite{Budzanowski10a}) that, away from the near-threshold region, the differential
cross section for $pp\to K^+ p\Lambda$ is essentially the same to
the left and right of the $\Sigma N$ threshold, in marked contrast even
to the phase-space behaviour." Inspection of the figure in Ref. \cite{Valdau11} shows that FSI was ignored when showing phase space dependence. This is in stark contrast to the experimental finding, as can be seen in Fig. \ref{Fig:Tof}, that including FSI makes the shape of the cross section underlying the peak rather flat. This is even more pronounced for the study in the lab. system.

%\clearpage
\section{Analysis of the peak structure}
\label{sec:Peak-structu-reactio-mechani}
Finally, Breit-Wigner distributions
\begin{equation}
\frac{{d^2 \sigma }}
{{d\Omega _{K,\pi } dm_{\Lambda p} }} =
 \frac{1}{\pi} \frac{d\sigma}{d\Omega _{K,\pi }}
\frac{\Gamma/2}{(m(\Lambda p)-m_{\text{0}})^2-(\Gamma/2)^2}
\label{equ:resonance}
\end{equation}
were fitted to the different yields.
Fit parameters were the width $\Gamma$ and $m_{0}$, the centroid of the peak.
The production cross section ${d\sigma}/{d\Omega _{K,\pi }}$ is also fitted.
\begin{figure}
\begin{center}
\includegraphics[height=0.60\textheight]{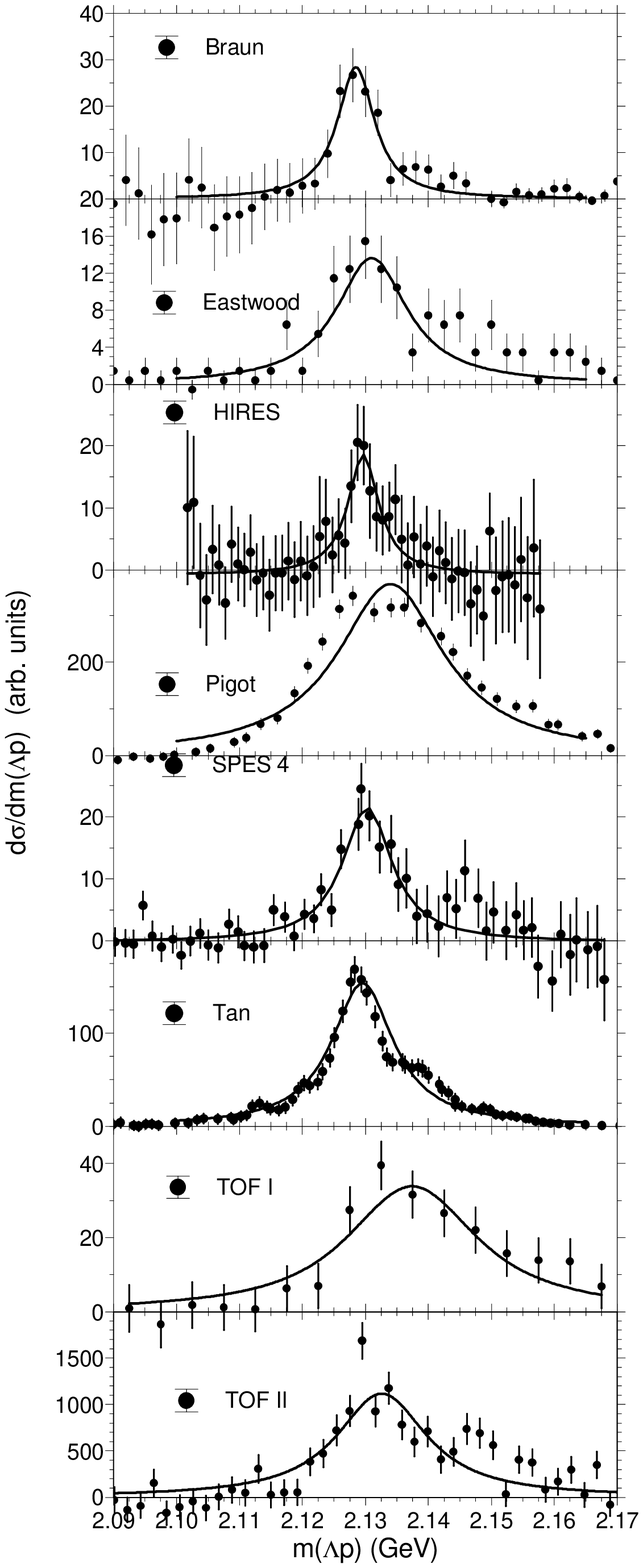}
\caption{Data after subtraction of cross section with smooth behavior
(for the $K^-d\to \pi^-\Lambda p$ reaction) or subtracted phase space distributions for $pp\to K^+\Lambda p$ and $pp\to K^+\Sigma^+n$ as discussed in previous sections. The data (full dots with error bars)  are from Braun \textit{et al.}
\cite{Braun77},  Eastwood \textit{et al.} \cite{Eastwood71},
Pigot \textit{et al.} \cite{Pigot85}, Tan \cite{Tan69}, SPES~4 \cite{Siebert94}, HIRES \cite{Budzanowski10a}, and TOF \cite{TOF_12_1}. The data from \cite{Braun77, Eastwood71, Pigot85, Tan69}
are from strangeness exchange (\ref{equ:K_induced}), the others
are from associated strangeness production (\ref{gat:reac3}).
Also shown are fits with a Breit-Wigner distributions (solid curves).}
\label{Fig:0ne-BW}
\end{center}
\end{figure}
In Fig. \ref{Fig:0ne-BW} these fits were compared to the data. The deduced centroids and widths are shown in Fig. \ref{Fig:0ne-BW}.
\begin{figure}[h]
\begin{center}
\includegraphics[width=0.6\textwidth]{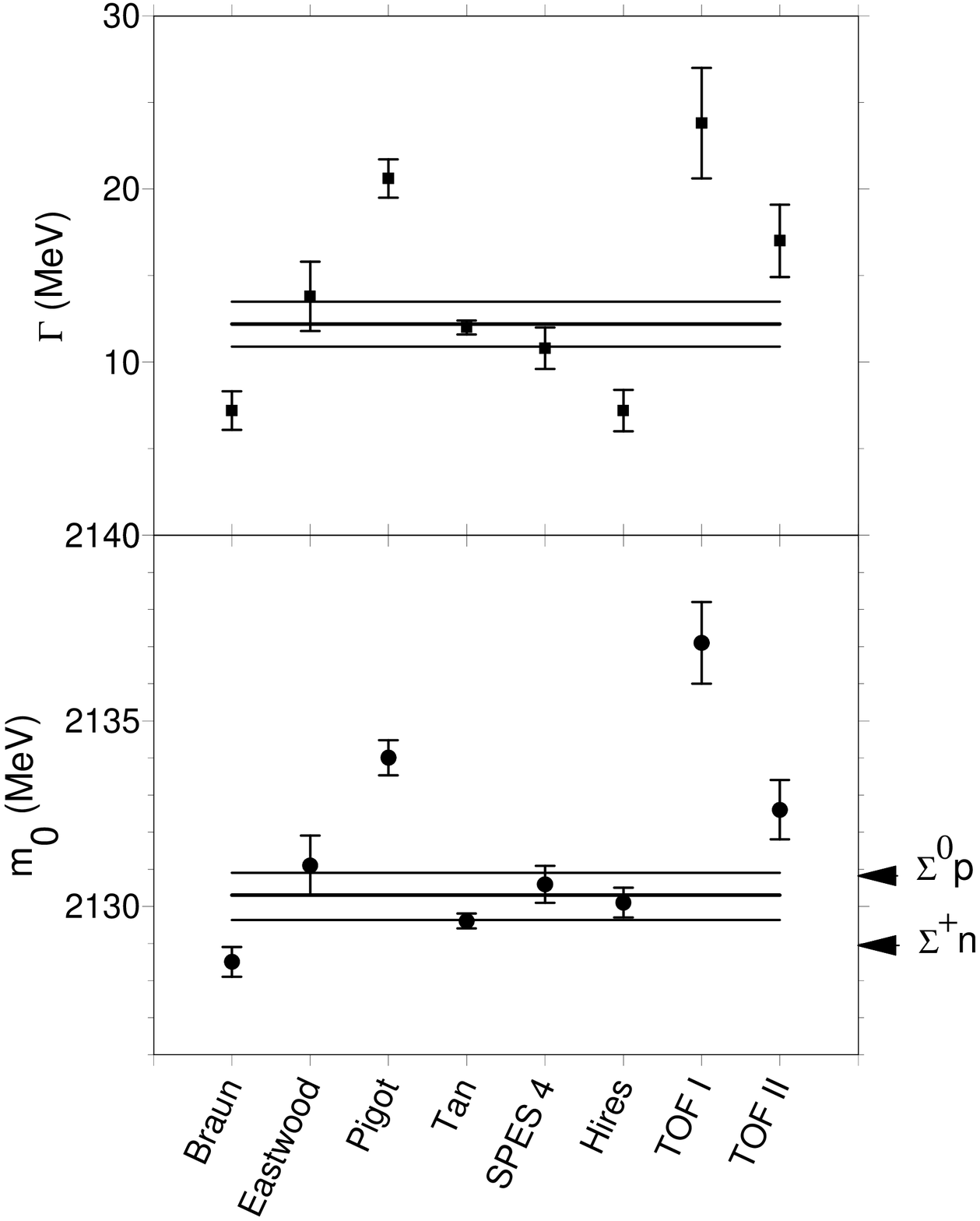}
\label{Fig:One-Lorentz}
\caption{The fitted centroid values (lower panel) and widthes (upper panel) from the fits with one Breit-Wigner distribution for the data sets from Braun \textit{et al.}  \cite{Braun77},  Eastwood \textit{et al.} \cite{Eastwood71}, Pigot \textit{et al.}
\cite{Pigot85}, Tan \cite{Tan69}, SPES4 \cite{Siebert94}, HIRES \cite{Budzanowski10a}, and TOF \cite{TOF_12_1}. The lines show the mean (thick line) and its variance (thin lines).
The arrows show the positions of the indicated thresholds.}
\end{center}
\end{figure}

The fits yield a mean for the centroids $m_{\text{0}}=2130.3\pm0.6$ MeV and for the widths $\Gamma=12.2\pm 1.3$ MeV. There is a large scattering among the values and this is reflected by large $\chi^2/dof$ values. Especially the results for the data set TOF~I are larger than the other values.
This data set shows an asymmetric peak with a centroid at $2.132 \pm 0.006$ GeV \cite{TOF_12_1}.
When a symmetric peak is fitted the centroid moves further up to $2.1376$ GeV. In particular, this is also in disagreement with the values extracted for the recent and more precise
TOF~II data. We, therefore, exclude these data from further analysis.

From inspection of the fits one gets the impression that the peak centroids are, with the exception of the data from Braun \textit{et~al.} \cite{Braun77}, above the $\Sigma^+n$ threshold which is also indicated in Fig. \ref{Fig:0ne-BW}. By way of example, Tan's data yield a centroid at $2129.55\pm 0.18$ MeV and the peak width $\Gamma = 12.0\pm 0.46$ MeV. The centroid is higher than in the analysis of Tan who reported $2128.7\pm 0.2$ MeV. Also the width is much larger than the $7.0\pm 0.6$ MeV reported by Tan. The reason is that he had fitted two Breit-Wigner functions to also account for a higher mass shoulder. We proceed by also fitting two Breit-Wigner distributions to all data. Then the peaks become smaller and there is good agreement between Tan's result and the one here. The agreement within errors shows the adequacy of the present method.

\begin{figure}[h]
\begin{center}
\includegraphics[height=0.60\textheight]{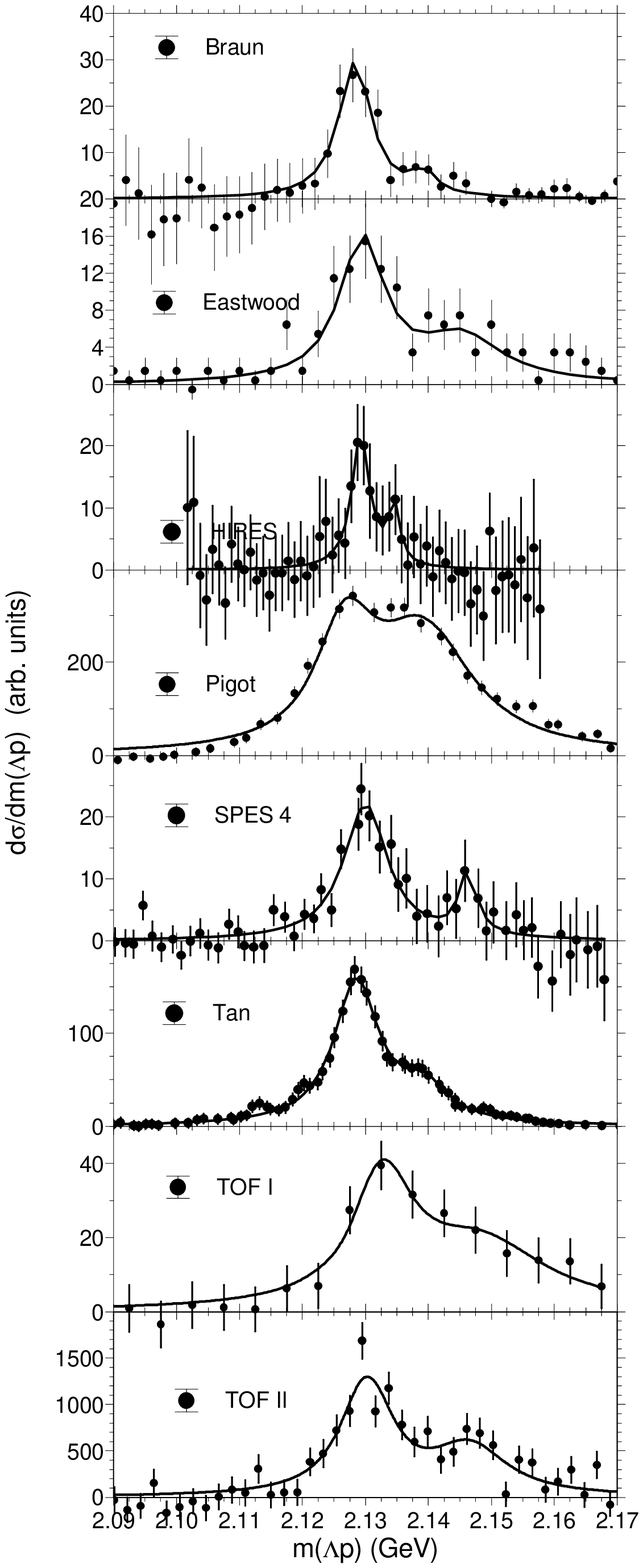}
\caption{Same as Fig. \ref{Fig:0ne-BW} but for two Breit-Wigner distributions fitted to each data set.}
\label{Fig:all-peaks}
\end{center}
\end{figure}

One feature of the data which has been also
ignored so far is a shoulder in the peak distribution at higher
missing masses. In order to take this shoulder into account we have
fitted incoherently two Breit-Wigner forms to the data. These fits are - on a $\chi^2$ criterion - slightly superior to those with only one Breit-Wigner. A fit on amplitude level did not indicate an interference. The peak
region after subtraction of a smooth cross section fraction is shown
for all data considered in Fig. \ref{Fig:all-peaks}.

Also shown are the fits with two Breit-Wigner functions. They account also for the higher mass range which was not in the case with a single Breit-Wigner (compare Fig. \ref{Fig:0ne-BW}). The fits with two Breit-Wigner functions are also superior to those with only one Breit-Wigner on a $\chi^2$ basis.
\begin{table}[h]
  \centering
  \caption{The reduced $\chi^2$-values obtained in the fits wit one Breit-Wigner (1 BW) and two Breit-Wigner forms (2 BW). }\label{Tab:Chi}
  \begin{tabular}{lcc}
\hline
\multicolumn{1}{l}{Ref.} & $\chi^2$/dof (1 BW) & \multicolumn{1}{l}{$\chi^2$/dof (2 BW)} \\
\hline
Braun & 0.64 & 0.63 \\
Eastwood & 0.92 & 0.58 \\
Pigot & 2.93 & 1.62 \\
Tan & 0.97 & 0.41 \\
SPES4 & 1.5 & 1.2 \\
HIRES & 0.42 & 0.32 \\
TOF II & 1.63 & 1.0 \\ \hline
\end{tabular}
\end{table}
This can be seen from the compilation in Table \ref{Tab:Chi}. The many values with $\chi^2$/dof$<$1 indicate that the authors have been very conservative in estimating the errors.

Obviously, there is a clear peak present in all data sets independent
of the reaction. The shoulder visible for larger invariant masses appears
also in all data sets. However, the latter looks somewhat different for
the two reactions. The fit with two Breit-Wigner functions accounts well for the data sets in all cases.

\begin{figure}[h]
\begin{center}
\includegraphics[width=0.45\textwidth]{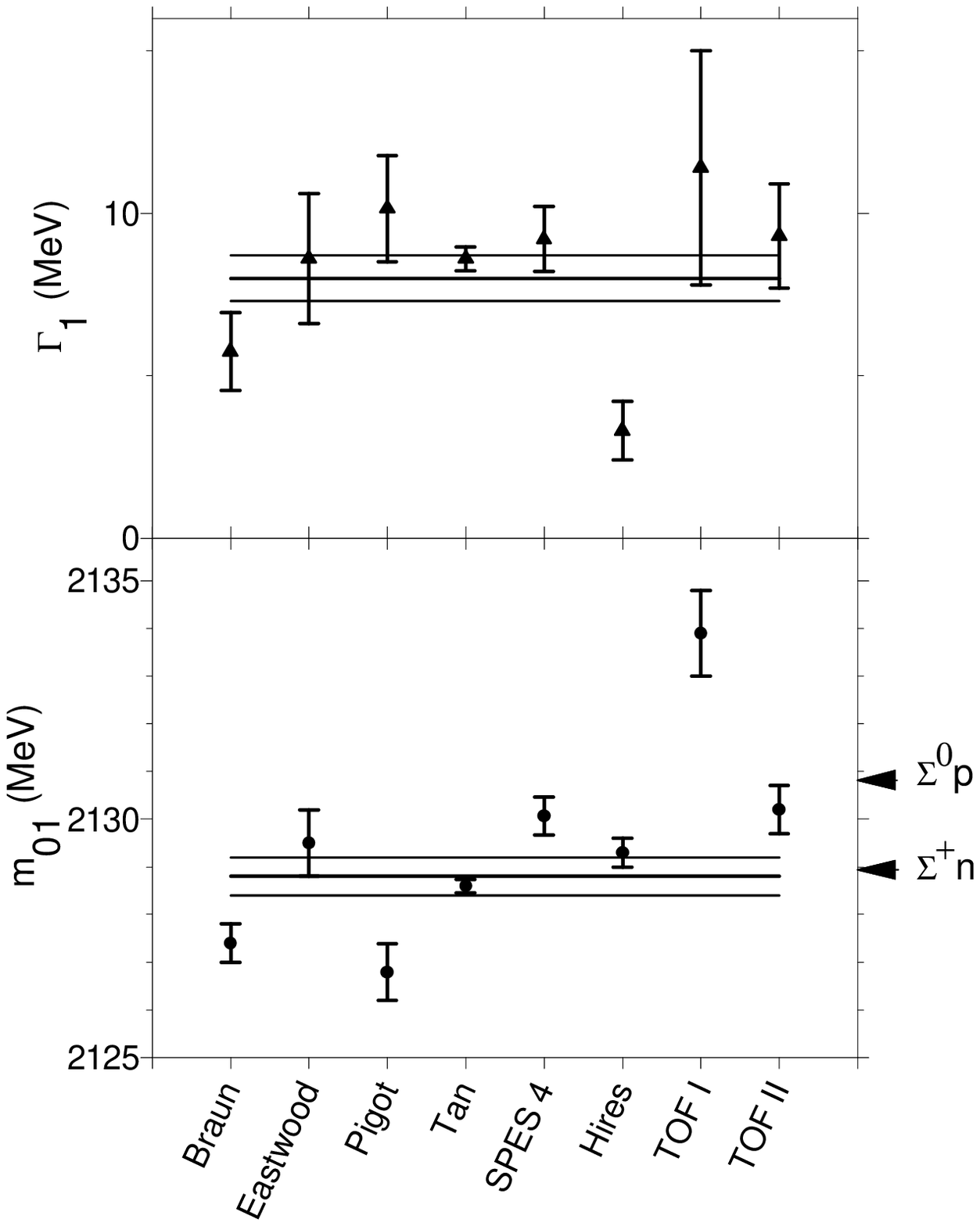}
\includegraphics[width=0.45\textwidth]{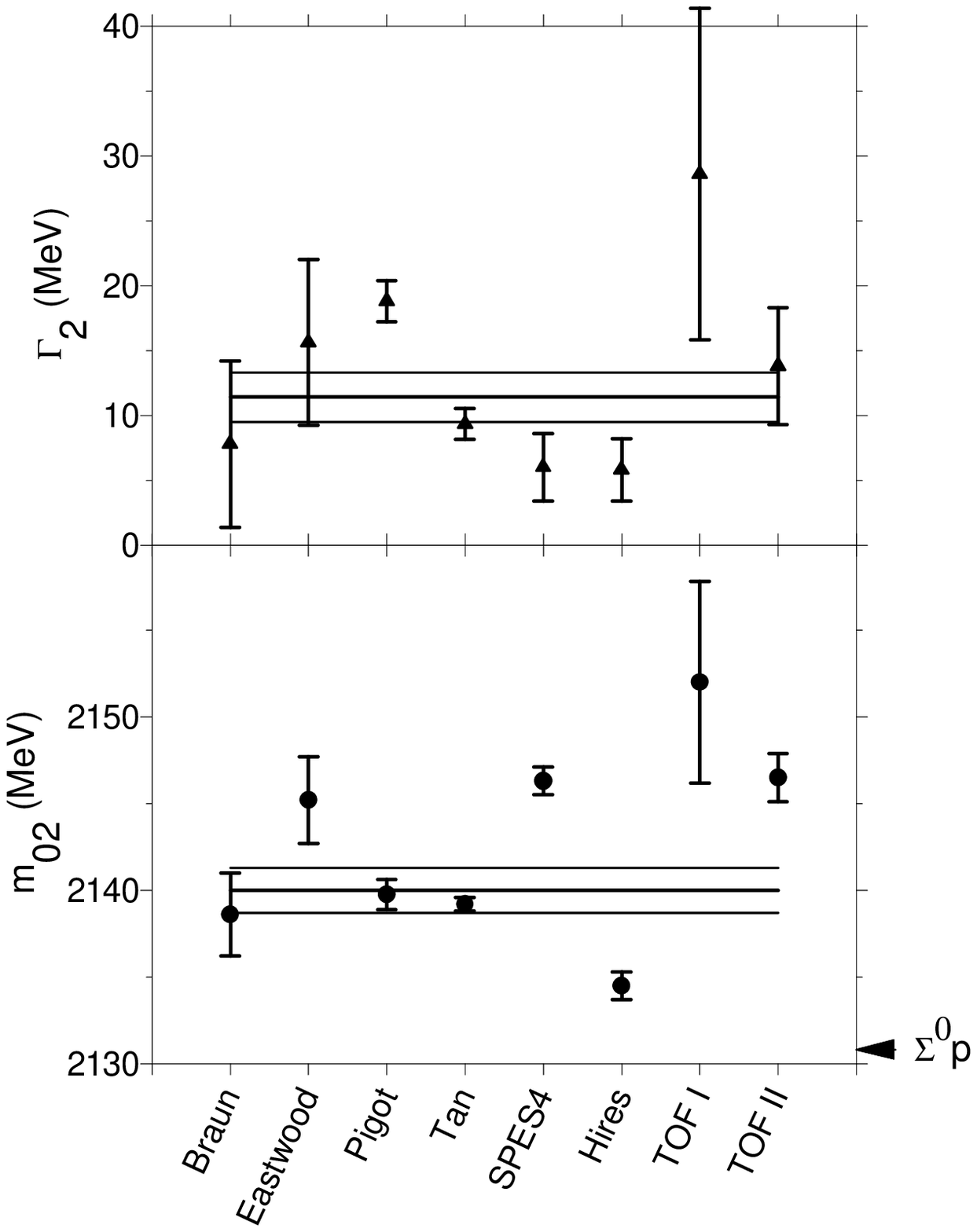}
\caption{Same as Fig. \ref{Fig:One-Lorentz}. Left panel: The Breit-Wigner parameters for the low mass structure for the data sets from Braun \textit{et al.}  \cite{Braun77},  Eastwood \textit{et al.} \cite{Eastwood71}, Pigot \textit{et al.}
\cite{Pigot85}, Tan \cite{Tan69}, SPES4 \cite{Siebert94}, HIRES \cite{Budzanowski10a}, and TOF \cite{TOF_12_1}. The lines show the mean (thick line) and its variance (thin lines).
The arrows show the positions of the indicated thresholds. Right panel; same as left panel but for the high mass structure. Note the wider scale on the right side compared to the left side.}
\label{Fig:peak_1}
\end{center}
\end{figure}

The fit parameters for the two Breit-Wigner forms are visualized in Fig.
\ref{Fig:peak_1}. Again the values obtained for the data set TOF~I are larger than all other results. The mean values and variances are also shown and
given in Table \ref{Table}. The values for the $\chi^2$/dof indicate
rather poor fits. For the lower-mass peak the centroids $m_{\text{01}}$
scatter around the result for Tan's data. In all cases the fit
parameters scatter. A plausible reason for this could be
uncertainties in the beam momenta. However, then one would expect both
centroids for a given measurement to be above or below the mean value.
This is not the case. It should be mentioned that the second
 higher mass structure for the $pp\to K^+\Lambda p$ reaction is smaller than for the $K^-d\to\pi^-\Lambda p$ reaction. This may point to differences in the reaction mechanism for the
two processes, which has to be expected anyway.

\begin{table}[h]
\centering \caption{The mean of the fitted Breit-Wigner distribution
parameters excluding the TOF~I data set. The centroids are denoted with $m_{\text{0i}}$ and the widths
with $\Gamma_i$.}\label{Table}
\begin{tabular}{ccc}
\hline
variable & fitted value (MeV) & $\chi^2/$dof \\ \hline
$m_{\text{01}}$ & 2128.7$\pm$0.3 & 8   \\
$\Gamma_1$ & 8.0$\pm$0.7 & 6   \\
$m_{\text{02}}$ & 2140.0$\pm$1.4 & 23  \\
$\Gamma_2$ & 11.3$\pm$ 2.0 & 6  \\
\hline
\end{tabular}
\end{table}

\subsection{The lower-mass peak}\label{sub:lower-mass-peak}

We are now left with the question of what the origin of the
structure is. We have shown above that the lower-mass peak position
coincides with the $\Sigma^+n$ threshold. This would hint to a
genuine cusp effect \cite{Miyagawa99}. However, the question arises
to what precision the experimental missing masses are known.
Possible shifts in the missing mass distributions in the bubble
chamber measurements were minimized by studying the chamber magnetic
field, energy range relation or kinematic fits to reference
reactions. In the case of the kaon beam applied by Pigot et al.  \cite{Pigot85} the incident
beam was carefully measured with the help of a magnetic
spectrometer. In the proton-proton experiments the question reduces
to the precision with which the accelerator beam momentum is known.
In lower energy experiments at COSY J\"{u}lich employing the
spectrograph Big Karl deviations of 0.11 to 0.5 MeV/c were found
\cite{Drochner98} and the COSY beam energy in later experiments was
corrected for these small deviations. Similarly for momenta around
1930 MeV/c a deviation of 2 MeV/c toward higher momenta was found
\cite{Betigeri99a}. Such a deviation by 2 MeV/c would amount to a
shift up of 0.7 MeV for the HIRES experiment. However, no correction
was found to be necessary when taking the $\Lambda p$ threshold as
benchmark \cite{Budzanowski10}. Note that the demanded agreement between
the HIRES and the SPES4 data required a shift of the latter to the low
energy side closer to the nominal accelerator beam momentum. From
these findings we conclude that the energy scale in the data used
here is reliable and therefore the lower-mass peak is at the
$\Sigma^+n$ threshold within two standard deviations of the
experimental uncertainty. Thus, there is no
evidence for a resonance in the $\Lambda p$ system below the
$\Sigma N$ threshold which in turn restricts the possibility of
the existence of a strangeness $S=-1$ counterpart of the
deuteron \cite{Torres86}.
Of course, nothing can be said for the region within that uncertainty
so that a quasi-bound deuteron-like state extremely close to the
$\Sigma^+n$ threshold cannot be excluded at present. We will study this possibility further.

\subsection{Flatt\'e approach}
\label{sec:Flatte}

At this stage let us emphasize that the Breit-Wigner distributions are used here solely as a tool
to determine the peak positions of the structures seen in the various experiments and to obtain
values for the widths that can be compared with each other. They are not meant as a physical
interpretation of the data. There are other parameterizations that are considered to be more
adequate for amplitudes
near a threshold like the one proposed by Flatt\'e \cite{Flatte76}, specifically, if one wants to
determine also the pole positions. However, those forms usually involve also more free parameters.
Furthermore, the finite resolution of the experiment requires that the yield should be folded with
Gaussians so that the functional form in the peak region would be identical anyway
(see Braun et al. \cite{Braun77}).

In any case, we performed also exploratory calculations with the relativistic formulae of
Flatt\'{e}~\cite{Flatte76}.
That the peak in $\Lambda p$ occurs just at the $\Sigma^+n$ threshold
is no accident, if the production of the two hyperon-nucleon states
are viewed as occurring via s-wave hyperon-nucleon scattering. As Flatt\'{e}
has pointed out then the drop on the high-mass side may be interpreted as the
effect of the opening of the $\Sigma^+n$ channel operating through unitarity.
But the imposition of analyticity requires that
the presence of the $\Sigma^+n$ channel must be
felt below threshold as well, thus creating the rapid
decrease in the $\Lambda p$ cross section on the low-mass side
of the peak.
The differential cross section around the $\Sigma^+n$ threshold is then
\begin{gather}\label{gat:above}
\frac{d\sigma_\nu}{dm}=C\left|\frac{m_r\sqrt{\Gamma_0\Gamma_\nu}}
{m_r^2-m^2-im_r(\Gamma_1+\Gamma_2)}\right|^2
\end{gather}
with $\nu=1$ denoting $\Lambda p\to \Lambda p$
and $\nu=2$ denoting $\Sigma^+ n\to \Lambda p$. The partial widths are
\begin{equation}
\Gamma_1=\Gamma(\Lambda p\to \Lambda p)=g_1k_1
\end{equation}
with $g_1$ an (effective) coupling constant squared~\cite{Flatte76} and $k_1$ the $\Lambda p$ c.m.
momentum so that $m = \sqrt{m_\Lambda^2+k_1^2}+\sqrt{m_p^2+k_1^2}$.
$C$ denotes an overall normalization constant which includes the
production cross section. $m_r$ is the mass of the resonance and $\Gamma_0$
the partial width into the incoming channel. It is usually assumed to be weakly
momentum dependent and thus can be assumed to be constant~\cite{Flatte76}.
For the other channel we have
\begin{equation}
\Gamma_2=\Gamma(\Sigma^+n\to \Lambda p)=g_2k_2
\end{equation}
with $k_2$ the $\Sigma^+n$ c.m. momentum. Above the $\Sigma^+n$ threshold $k_2$
is real but below the threshold it becomes purely imaginary so that
\begin{equation}
\Gamma_2=\Gamma(\Sigma^+n\to \Lambda p)=g_2\kappa_2
\end{equation}
with $\kappa_2=i|k_2|$. In this region the cross section is given by
\begin{equation}\label{equ:below}
\frac{d\sigma_1}{dm} = C \frac{m_r^2\Gamma_0\Gamma_1}{(m_r^2-m^2+m_r\Gamma_2)^2 +m_r^2\Gamma_1^2}
\end{equation}
while $d\sigma_2/dm=0$.

In the actual calculations we assumed that $\Gamma_0=g_1k_1(m_r)$. In practice, $\Gamma_0$ can be
absorbed into the overall normalization factor $C$.
It turned out that the fits are not sensitive to the resonance parameters,
something that was also seen already by Braun et al.~\cite{Braun77} when analyzing
their own data. Thus, it seems difficult to learn more on the pole position.
Therefore, we kept those parameters fixed to reproduce a peak at the ($\Sigma^+ n$) threshold and fitted only the other parameters (couplings and an overall normalization). In addition we assumed a Gaussian smearing with a fitted width to take into account the finite resolution of the different experiments. Although there is a large variation in the obtained values, the ratios of the coupling constants are better defined. Indeed, this is just a reflection of the scaling properties of the Flatt\'e parametrization discussed in Ref.~\cite{Baru05}. It is worthwhile to mention that the final curves due to the folding with the Gaussian resolution function are quite similar to Breit-Wigner functions.

The standard Flatt\'e parametrization Eq.~(\ref{gat:above}) produces only a single
peak. It cannot generate the secondary peak (or shoulder) seen in the data.
In order to account for this component we have coherently added a Breit-Wigner
function. However, in the actual fit procedure the centroid of the
Breit-Wigner is always moved to the maximum while simultaneously the
Flatt\'e part is reduced.
The fits now show a wider distribution but with almost no interference.
It should be stressed that the assumed Flatt\'e formalism is only for a
coupled two-channel system. In the discussed $\Lambda N - \Sigma N$ system
there is also a tensor coupling between the $^3S_1$ and $^3D_1$ partial
waves so that effectively, one deals with a system of four coupled channels
and one would need to use a suitably generalized form for the Flatt\'e
parameterizations \cite{Badalyan82}.

\subsection{The higher-mass structure}\label{sub:higher-Mass-structu}
In the case of the second structure, which in most of the data sets looks
like a shoulder, there is no near-by threshold.
Therefore one possible explanation is indeed that there could be a genuine
resonance in the $\Lambda p$ system.
However, such structures could be also artefacts of the cuts applied to the data from the
$K^-d\to \pi^-\Lambda p$ reaction
in order to exclude events from processes with the proton being a spectator.
The momentum distribution of the nucleons in the deuteron is asymmetric with respect to its mean, say a Hulth\'{e}n function
with a maximum at 50 MeV/c. Energy cuts on the proton spectra in this order of magnitude were
applied as discussed in Sect.~\ref{sec:reactio-K-}. Deloff \cite{Deloff89a} showed that such
cuts on the proton energy as well as on the proton emission angle affects the spectrum for
$m(\Lambda p)<2.1$ GeV. In order to study whether such cuts can produce structures in the spectra
at higher mass values we perform Monte Carlo (MC) calculations. The quasi-free production is ignored.
In addition to phase space the lower-mass peak is represented by a Breit-Wigner distribution (\ref{equ:resonance})
with $m_\text{0}=m_{\text{01}}$ and $\Gamma=\Gamma_1$ the corresponding parameters from Table \ref{Table}.
\begin{figure}[h]
\begin{center}
\includegraphics[width=0.6\textwidth]{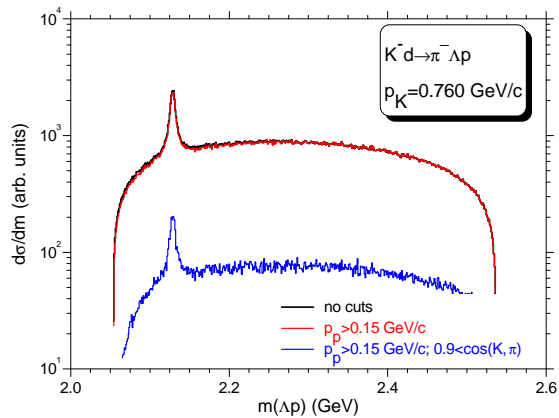}
\caption{Monte Carlo calculations for the strangeness exchange reaction $K^-d\to \pi^- \Lambda p$ at a beam momentum of 0.760 GeV/c. A peak with Breit-Wigner shape for the peak with mass $m_01$ and width $\Gamma_1$ from Table \ref{Table} is added to phase space distribution. The spectrum without cuts is shown in black, with a cut on the proton momentum is shown in red and the one with an addition cut on the pion emission angle is shown in blue.}
\label{Fig:760}
\end{center}
\end{figure}
A~beam momentum of 760 MeV/c was chosen, which corresponds to the case of Ref. \cite{Braun77}.
The $\Lambda p$ spectrum obtained in this way is shown in Fig. \ref{Fig:760} as black curve.
In a next step the requirement  $p_p\geq 150$ MeV/c was chosen (red curve). The effect on
the spectrum is minor. We then introduced the requirement $\cos(K^-,\pi^-)>0.9$. This reduces
the height of the spectrum severely with, however, no change in the spectral shape (blue curve).
This is not surprising since below the peak there is phase space distribution and hence isotropic emission.

Another effect studied to inspect variations of the spectral shape are crossed-channel resonances.
The Dalitz plot shown by Braun \textit{et al.} \cite{Braun77} showed the excitation of the
$\Sigma(1385)$ clearly. This resonance at the beam momentum of $\approx 760$ MeV/c does not have
an overlap with the peak at $\approx 2129$ MeV.
\begin{figure}[h]
\begin{center}
\includegraphics[width=0.6\textwidth]{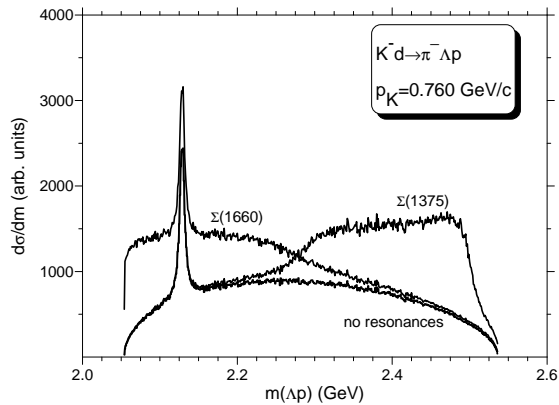}
\caption{MC calculation for the $\Lambda p$ spectrum for the reaction
$K^- d \to \pi^-\Lambda p$ for a kaon momentum of 0.76 GeV/c.
Influence of crossed-channel $\Sigma$ resonances.
Same description of the calculation as in Fig.~\ref{Fig:760}.
}
\label{Fig:Sigma_res_760}
\end{center}
\end{figure}
This is also visible in the $\Lambda p$ spectrum shown in Fig. \ref{Fig:Sigma_res_760}.
Resonance mass and width for that simulation are taken from the PDG \cite{PDG10}. A putative
$\Sigma(1480)$ resonance (referred to as $\Sigma(1480)$ bumps by the PDG \cite{PDG10})
is not visible in the projection of the Dalitz plot on the
$\pi^-\Lambda$ axis in \cite{Braun77}. Also such a
resonance would not lie below the peak at 2129 MeV. The PDG \cite{PDG10} suggests in a note
that there are two $\Sigma$ resonances around 1670 MeV:
a $P_{11}$ at $\approx1660$ MeV with $\Gamma\approx 100$ MeV
with three stars and a $D_{13}$ at $\approx1670$ MeV with width $\Gamma\approx 60$ MeV
with four stars.
In order to account for these resonances we have assumed a mass of 1660 MeV and a width of 100 MeV.
Although the Dalitz plot in \cite{Braun77} showed that the strengths of these resonances, if excited,
are smaller than for the $\Sigma(1385)$, we assumed the same production cross section, since we are
only interested in a possible change of the spectral shape. The peak area is clearly shifted up by
the introduction of a such resonance, cf. Fig. \ref{Fig:Sigma_res_760},
but the cross section below the peak remains smooth. Finally we
examine the influence of crossed-channel resonances on data taken at kaon beam momenta almost at rest.
This is shown in Fig. \ref{Fig:MC_1660}.
\begin{figure}[h]
\begin{center}
\includegraphics[width=0.6\textwidth]{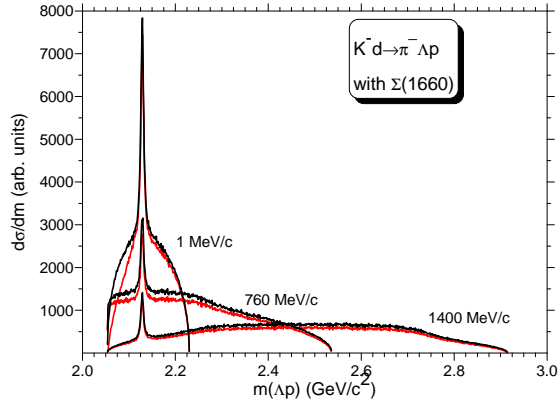}
\caption{Same as Fig. \ref{Fig:Sigma_res_760} but for only $\Sigma(1660)$. The kaon beam momenta are next to the appropriate curves. The upper curves (black) are without cuts while the lower ones (red) are with cuts.}
\label{Fig:MC_1660}
\end{center}
\end{figure}

A possible explanation for a shoulder at higher-mass was given in
Toker \textit{et al.} \cite{Toker81}. In that work a shoulder was observed
due to an interference between the ``direct`` $\Lambda$ production mechanism
and those involving the $\Sigma N \to \Lambda N$ transitions amplitude,
in case of their interaction that produces an inelastic virtual state. We should
add, however, that such a scenario was later on disputed by Deloff \cite{Deloff89a}.
Rather, he advocated an interpretation that resorted to the presence of a $P$-wave
resonance at 2140 MeV. In fact, enhanced $P$-wave contributions around the
$\Sigma N$ threshold are certainly possible. For example, an inspection of the
$\Lambda p\to \Lambda p$ partial waves of the J\"{u}lich $YN$ model, cf. Fig.~13 in
\cite{Haidenbauer05}, reveals the $^1P_1$-wave shows indeed some structure. As
a consequence, there is a shoulder in the elastic scattering cross section at a few
MeV above the $\Sigma N$ threshold, see the solid line in Fig. \ref{Fig:elastic}.
Its maximum is at 2139 MeV, when transformed to the $\Lambda p$ invariant mass,
which is in good agreement with the experimental findings
(see Table \ref{Table}). However, the width is certainly larger than the experimental
results.

The differences between the data for the two reactions are that the data from reaction $K^-d\to \pi^-\Lambda p$ are total cross sections whereas those for the $pp\to K^+\Lambda p$ reaction are differential cross sections. Hence in the latter case interferences are possible which are absent in the first case. We study the second peak in the case of the second reaction further. Here we concentrate the discussion on the HIRES case. We may assume the second peak structure as a statistical fluctuation. A fit with only one Breit-Wigner yields  $\chi^2$/dof= 22.4/56, while a fit with two Breit-Wigner functions yields 15/53.
At this point it should be mentioned that the large error bars are not due to poor statistics but due to systematic uncertainties like flight paths in the spectrometer and decays along this path as well as uncertainties in the acceptance. Therefore a good fit does not yield $\chi^2$/dof= 1 as is the case of statistically distributed data. In a further step we fit one Breit-Wigner by omitting the data in the range of the second peak. The remaining cross section was then fitted by a second Breit-Wigner.
\begin{figure}[h]
\begin{center}
\includegraphics[width=0.5\textwidth]{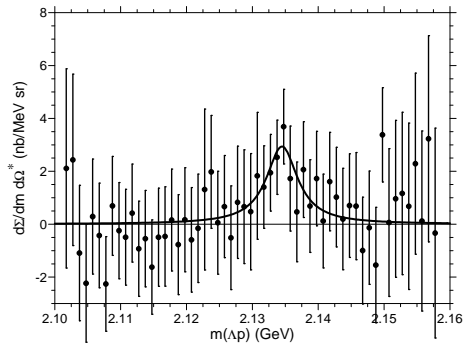}
\caption{The remaining cross section after subtracting the low mass peak from the HIRES data. The solid curve is a fit with a Breit-Wigner function.}
\label{Figh:Right}
\end{center}
\end{figure}
This is shown in Fig. \ref{Figh:Right} together with a fitted Breit-Wigner distribution. Note that the error bars contain the uncertainties from the different subtraction steps. The peak shows a significance of 4.0$\sigma$. This is smaller than the common practice of requiring 5$\sigma$ for a discovery.

\section{Summary and Discussion}
\label{sec:Discuss-Summary}

We have studied the $\Lambda p$ interaction in the vicinity of the
$\Sigma^+ n$ and $\Sigma^0 p$ thresholds. Experimental elastic transition
cross sections do not show any particular enhancement at the threshold for
the $\Lambda p\to \Sigma N$ scattering. However, the quality of the
existing data is poor and does not allow for reliable conclusions. While
the enhancement predicted by the recent J\"{u}lich OBE model \cite{Haidenbauer05}
is modest, it is much larger in the Nijmegen soft-core potentials
\cite{Maessen89,Rijken99} and in the leading-order $YN$ interaction obtained
within the chiral effective field approach~\cite{Polinder06}.

In the study of the $\Lambda p$ \fsi in case of the strangeness exchange
reaction $K^-d\to\pi^-\Lambda p$ the experiments made cuts of 75-150
MeV/c on the proton momenta, thus excluding real spectator protons.
After applying the cuts the peaks at the $\Sigma N$ thresholds from
different experiments became comparable with respect to their shape.
The influence of the momentum cuts as well as angular cuts were
studied in detail by Deloff~\cite{Deloff89a}.

The enhancement is also visible in high resolution experiments
employing the reaction $pp \to K^+\Lambda p$. A good agreement is
found with respect to the shape for those data from strangeness production Refs. \cite{Siebert94, Budzanowski10a} and \cite{TOF_12_1}.
When comparing the latter, where the data from Refs. \cite{Siebert94} and \cite{Budzanowski10a} are transformed into the c.m system, with those
from the reaction $K^-d\to\pi^-\Lambda p$, agreement is found for the
structure around the $\Sigma N$ threshold with cuts applied. This is
expected since in the $pp\to K^+\Lambda p$ reaction both nucleons participate
due to $\pi$ or $K$ exchange. However, it seems mandatory to augment the $pp\to  K^+\Lambda p$ data when new exclusive measurements with high resolution close to threshold become available.

The coincidence of the peak maximum with the $\Sigma^+n$ threshold
within the experimental errors suggests that the peak is most likely
due to a cusp effect induced by the strong coupling between the
$\Lambda N$ and $\Sigma N$ channels. Some of the
Nijmegen $YN$ potentials produce a cusp in the $^3S_1-^3D_1$
partial wave \cite{Maessen89,Rijken2010}.
The same is true for the recent J\"{u}lich model \cite{Haidenbauer05}.
Conclusive evidence for an unstable (deuteron-like) bound state in
the $\Sigma N$ (isospin 1/2) channel, as predicted by other $YN$
potentials in the literature \cite{Holzenkamp89,Rijken99},
would require a peak that is well separated from the (and below the)
$\Sigma^+n$ threshold.

In both reactions there is a second peak at invariant masses well above the $\Sigma^0 p$
threshold. A possible interpretation due to deformation of the lower-mass peak
by experimental cuts, crossed-channel resonances or interference effects can
be excluded.
Some $YN$ potentials like the recent J\"{u}lich model \cite{Haidenbauer05}
predict an enhanced $P$-wave contribution around the $\Sigma N$ threshold,
cf. the elastic $^1P_1$ phase shift in Fig.~13 of that reference. This leads to a
shoulder in the $\Lambda p$ cross section in the relevant energy region,
which at least qualitatively resembles the shoulder seen in the reactions
discussed in the present study.
But more exotic interpretations in form of a genuine $\Lambda p$ resonance
are not excluded. For example, it might be a $S=-1$  state as predicted by Aerts
and Dover \cite{Aerts84, Aerts85}.
However, their predicted $^1P_1$ state lies below the $\Sigma N$ threshold.
A recent review of $S=-1$ dibaryonic states can be found in Ref.
\cite{Gal11}.

Model calculations that study the enhancement seen in the $\Lambda p$ invariant
mass near the $\Sigma N$ are all based on $YN$ interactions that include the
coupling between the $\Lambda N$ and $\Sigma N$ channels for obvious reasons
\cite{Toker81,Torres86,Deloff89,Laget91}.
The investigations by Toker et al.~\cite{Toker81} and Torres et al.~\cite{Torres86}
deal with the reaction $K^- d \to \pi^- \Lambda p$ and are performed in the Faddeev formalism.
In Ref.~\cite{Toker81} different potential models are employed that produce an inelastic
virtual state (and thus a threshold cusp) but also an unstable bound state, where
in the latter case a resonance-like behavior in the $\Lambda p$ cross section is
generated slightly below and even above the threshold. No direct comparison of their
results with the $\Lambda p$ spectrum of Tan \cite{Tan69} is presented.  Torres et al. \cite{Torres86} achieve quantitative agreement with the data. They claim that
a virtual state near the $\Sigma N$ threshold is required for reproducing the data
but not an unstable bound state.
In the calculations for the associated strangeness production reaction (\ref{gat:reac3})
Deloff \cite{Deloff89} as well as Laget \cite{Laget91} use $YN$ interactions
that produce a virtual state, namely the models presented in Ref.~\cite{Deloff89a}
in the former case and the Nijmegen model ND \cite{Nagels77} in the latter.

New experiments with better statistics are required in order to discriminate between the different
scenarios. $\Lambda$ hyperons at low momenta could be produced with high intensity at the J-PARC
facility in Japan and then used for pertinent experiments. Also the strangeness exchange reaction
could be studied with the E31 setup again at J-PARC \cite{Noumi}.
With regard to the $pp\to K^+\Lambda p$ reaction exclusive data with full acceptance
but even higher resolution than the existing ones would be highly desirable.

%%%%%%%%%%%%%%%%%%%%%%%%%%%%%%%%%%%%%%%%%%%%%%%%%%%%%%%%%%%%%%%%%%%%%
%
\section*{Acknowledgements}\label{sec:Acknowl}
Two of us (H.M. and J.A.N.) acknowledge research exchange grants
from DAAD (50740781) and the Academy of Finland (139512). We are grateful to M. R\"{o}der for supplying the TOF~II data set.

%
%%%%%%%%%%%%%%%%%%%%%%%%%%%%%%%%%%%%%%%%%%%%%%%%%%%%%%%%%%%%%%%%%%%%%
%%%%%%%%%%%%%%%%%%%%%%%%%%%%%%%%%%%%%%%%%%%%%%%%
% Begin References
%%%GATHER{c:\Program Files\MiKTeX 2.9\bibtex\bib\eigene\NN_Interaction.bib}
%

%%\bibliographystyle{elsart-num}
%%\bibliographystyle{My}
%%\bibliography{NN_Interaction}

%%%%%%%%%%%%%%%%%%%%%%%%%%%%%%%%%%%%%%%%%%%%%%%%%
% End References
%%%%%%%%%%%%%%%%%%%%%%%%%%%%%%%%%%%%%%%%%%%%%%%%%
\end{document}